\documentclass[onecolumn,showpacs,11pt,nofootinbib]{revtex4-2}
\usepackage{axodraw}
\usepackage{graphicx}
\usepackage{dcolumn}
\usepackage{bm}
\usepackage{color}
\usepackage{eurosym}
\usepackage{multirow}
\usepackage{amsfonts}
\usepackage{amsmath}
\usepackage{hyperref}
\usepackage{amssymb}
\usepackage[english]{babel}
\usepackage{graphicx}
\usepackage{epsfig}
\usepackage{subfigure}
\usepackage{unicode}
\begin{document}
\newcommand{\hs}{\hspace*{0.2cm}}
\newcommand{\hsp}{\hspace*{0.5cm}}
\newcommand{\vs}{\vspace*{0.5cm}}
\newcommand{\be}{\begin{equation}}
\newcommand{\ee}{\end{equation}}
\newcommand{\bea}{\begin{eqnarray}}
\newcommand{\eea}{\end{eqnarray}}
\newcommand{\ben}{\begin{enumerate}}
\newcommand{\een}{\end{enumerate}}
\newcommand{\bde}{\begin{widetext}}
\newcommand{\ede}{\end{widetext}}
\newcommand{\nn}{\nonumber}
\newcommand{\crn}{\nonumber \\}
\newcommand{\Tr}{\mathrm{Tr}}
\newcommand{\non}{\nonumber}
\newcommand{\noi}{\noindent}
\newcommand{\al}{\alpha}
\newcommand{\la}{\lambda}
\newcommand{\bet}{\beta}
\newcommand{\ga}{\gamma}
\newcommand{\va}{\varphi}
\newcommand{\om}{\omega}
\newcommand{\pa}{\partial}
\newcommand{\+}{\dagger}
\newcommand{\fr}{\frac}
\newcommand{\sq}{\sqrt}
\newcommand{\bc}{\begin{center}}
\newcommand{\ec}{\end{center}}
\newcommand{\Ga}{\Gamma}
\newcommand{\de}{\delta}
\newcommand{\De}{\Delta}
\newcommand{\ep}{\epsilon}
\newcommand{\varep}{\varepsilon}
\newcommand{\ka}{\kappa}
\newcommand{\La}{\Lambda}
\newcommand{\si}{\sigma}
\newcommand{\Si}{\Sigma}
\newcommand{\ta}{\tau}
\newcommand{\up}{\upsilon}
\newcommand{\Up}{\Upsilon}
\newcommand{\ze}{\zeta}
\newcommand{\ps}{\psi}
\newcommand{\Ps}{\Psi}
\newcommand{\ph}{\phi}
\newcommand{\vph}{\varphi}
\newcommand{\Ph}{\Phi}
\newcommand{\Om}{\Omega}
\newcommand{\Red}[1]{{\color{red}#1}}
\newcommand{\Cyan}[1]{{\color{cyan}#1}}
\newcommand{\Revised}[1]{{\color{blue}#1}}
\newcommand{\Vien}[1]{{\color{blue}#1}}
\newcommand{\Blue}[1]{{\color{blue}#1}}
\newcommand{\Green }[1]{{\color{green}#1}}
\title{Neutrino phenomenology and keV dark matter in 2HDM with $A_4$ symmetry}
\author{V. V. Vien$^{a}$}
\email{vvvien@ttn.edu.vn}
\affiliation{$^{a}$Department of Physics, Tay Nguyen University, 
Daklak province, Vietnam.}
\begin{abstract}
We propose a  minimal extended seesaw scheme based on the discrete symmetry $A_4\times Z_4\times Z_2\times Z_8$ which can successfully address neutrino phenomenology and keV sterile neutrino dark matter. The lepton mass hierarchy is naturally achieved. Active neutrino mixing angles can reached the best-fit points with the predictive Dirac CP violation phase. The active-sterile mixing matrix elements are small enough to access the observed cosmological dark matter abundance constraint with keV sterile neutrino dark matter. The effective neutrino masses are predicted to be in the ranges of the recent experimental limits.
\end{abstract}
\date{\today}
\maketitle
\section{Introduction}
Although the mass hierarchy, the absolute neutrino mass and the Dirac CP-violating phase are still unknown, two squared neutrino mass differences and neutrino mixing angles have been measured with high accuracy \cite{Salas2020}. However, lepton mass hierarchy problem is one of the key subject in the elementary particle physics which cannot be explained within the framework of the standard model (SM). A possible solution to the fermion mass hierarchy problem is to introduce a new family symmetry acting between generations \cite{Raby2000,ALVEPJPC16,VienCJP22,VienJPG22,Vienmpla22D27}. Furthermore, one interesting anomaly derived from the experimental data of LSND \cite{LSND} and MiniBooNE \cite{MiniBooNE} that cannot be explained by the three-neutrino scenario but could be solved by adding at least one neutrino, motivating to extended scenarios with the existence of additional neutrinos which called sterile neutrinos since they mix with active neutrinos but do not participate in the SM gauge interactions. Besides, even though the strong evidences for the existence of dark
matter (DM) \cite{Rubin70,Clowe06,Cai18}, its properties is still an open question. It has been 
proved that   \cite{Gariazzojpg15,adhikari2017white,Boyarsky2019,Mertensjpg19,Akerepjc23} a keV-scale sterile neutrino is viable candidate for
warm DM. Planck data \cite{Planck2015} implies that $26.8 \%$ of the total energy density of
the Universe and the dark matter abundance is given by
\bea
\Omega_{\mathrm{D}} h^2\in (0.117, 0.120). \label{DMconstrain1}
\eea
The most recent result on the dark matter abundance from \cite{Planck2018} sets a new limit
\bea
\Omega_{\mathrm{D}} h^2\in (0.119, 0.121). \label{DMconstrain2}
\eea
Recently remarkable result by the STEREO experiment strongly rejected the hypothesis of the existence of a sterile neutrino with a few-eV-mass \cite{STEREO23}. However, sterile neutrinos with heavier masses such as keV scale and mixing very weakly with the active ones could exist but STEREO has not been sensitive to them or the mixing would be too weak to explain the reactor anomaly. 

The outstanding feature of discrete symmetries is that they can give a satisfactorily interpretation of
the neutrino oscillation data. The neutrino phenomenology and sterile neutrino dark matter with $A_4$ symmetry have been considered in Refs. \cite{Das2016,Das2020,Mishra2020,VienA4JPG22} which is different from our current work with significantly different scalars and for the following basic properties:
\begin{itemize}
  \item [$\bullet$] Refs. \cite{Das2016} based on symmetry $SU(2)_L\times U(1)_Y\times A_4\times Z_2\times Z_3$ with five additional fermion singlets, four Higgs doublets, four scalar singlets and one scalar triplet in the frame work of inverse and type II seesaw mechanism with  GeV-scale dark matter in which only inverted neutrino mass ordering is satisfied while the charged lepton mass hierarchy is not natural and active-sterile mixing angles are not mentioned.
  \item [$\bullet$] Ref. \cite{Das2020} based on symmetry $SU(2)_L\times U(1)_Y\times U(1)^'_{Y^'}\times A_4\times Z_4\times Z_3$ with two Higgs doublets and up to fifteen scalar singlets for NO (sixteen scalar singlets for IO) in the frame work of MES framework with  keV-scale dark matter in which the charged lepton mass hierarchy is not natural and the active-sterile mixing angles are not explicitly mentioned.
  \item [$\bullet$] Ref. \cite{Mishra2020} based on symmetry $SU(2)_L\times U(1)_Y\times U(1)_{B-L}\times A_4\times Z_3\times Z_2$ with one Higgs doublet and ten scalar singlets in 3 + 1 mixing scheme with GeV-scale dark matter in which the charged lepton mass hierarchy is not natural\footnote{In Ref. \cite{Mishra2020} the Yukawa couplings differ about five orders of magnitude from $y_{e}\sim \mathcal{O} (10^{-5})$ to $y_{1}\approx y_\chi \sim \mathcal{O} (1)$. In order to give a natural explanation to the charged lepton mass hierarchy, the Yukawa couplings in the charged-lepton differ about four orders of magnitude from $y_{e}\sim \mathcal{O} (10^{-5})$ to $y_{\tau}\approx y_\chi \sim \mathcal{O} (10^{-1})$ . } and the neutrino masses are
generated by the help of up to seven-dimension term and the active-sterile mixing elementes are predicted to be $U_{14}=U_{24}=U_{34}$ which don't seem natural.
\item [$\bullet$] Ref. \cite{VienA4JPG22} based on symmetry $SU(2)_L\times U(1)_Y\times U(1)_{B-L}\times A_4\times Z_3\times Z_4$ with one Higgs doublet and eleven scalar singlets in MES scheme with sterile neutrino mass in eV-scale and the sterile-active neutrino mixing elements $|U_{i4}|^2 \sim (10^{-3}, 10^{-2})$ that cannot be a WDM candidate.
\end{itemize}

In the present study, we propose a minimal extended seesaw (MES) \cite{Barry2011v,VienS3EPJC21v,VLAS4PTEP22v} with the discrete symmetry $A_4\times Z_4\times Z_2\times Z_8$ in the framework of the two Higgs doublets model with two $A_4$ triplet and two $A_4$ singlet flavon fields to explain neutrino phenomenology and keV sterile neutrino dark matter in both normal and inverted hierarchies.

The remaining part of this study is as follows. The description of the model is given in section \ref{model}. Its implications in neutrino phenomenology is given in \ref{neutrinomixing}. Section \ref{DMA4} is intended for the dark matter phenomenology. The numerical analysis is presented in section \ref{NR}. Section \ref{conclusion} contains some conclusions. 

\section{\label{model}The model}
We construct a MES model in which the SM model is supplemented by one non-Abelian discrete symmetry $A_4$ and three Abelian symmetries $Z_2, Z_4$ and $Z_8$. Three right-handed neutrinos $\nu_{1,2,3R}$ and one sterile neutrino $\nu_{s}$ are added to the SM. On the other hand, one doublet $H^'$ and eight singlets ($\phi_l, \phi_\nu, \chi, \rho$) are added
to the SM, i.e., the considered model contains two $SU(2)_L$ doublets\footnote{In the Two-Higgs doublet model (2HDM), two $SU(2)_L$ doublets $H$ and $H^'$ contain eight fields in which three of them (Goldstone bosons) are eaten to give mass to the gauge bosons $W^{\pm}$ and $Z^0$. The remaining five are physical Higgs states including two neutral scalars, one pseudoscalar and a charged pair. The masses of the gauge particles and the Higgs particles are in consistent with the experimental constraints which has been presented in various works (see, for instance \cite{Branco12, Wang23}, for a review of the 2HDM).}. In this model, three left-handed leptons ($\psi_L\equiv \psi_{1,2,3L}$) are assigned in $\underline{3}$ under $A_4$ symmetry while the first right-handed charged lepton $l_{1R}$, three right-handed neutrinos, and the sterile neutrino are assigned in
 $\underline{1}$ and the last two right-handed charged leptons ($l_{2R}, l_{3R}$) are assigned as $\underline{1}^{''}$ under $A_4$. The particle content of the model, under the symmetry $\mathcal{G}=SU(2)_L\times U(1)_Y\times A_4\times Z_4\times Z_2\times Z_8$, are summarized in Table \ref{particlecont}.

The charged-lepton masses can be obtained via the couplings of
$\bar{\psi}_{L} l_{1,2,3 R}$ to scalars in which, under $\mathcal{G}$  symmetry, $\overline{\psi}_{L} l_{1R}\sim (\mathbf{2}, -1/2, \underline{3}, -i, -1, \om^{4})$ and $\overline{\psi}_{L} l_{2,3R}\sim (\mathbf{2}, -1/2, \underline{3}, -i, 1, \om^{5})$, i.e., the scalar doublets which respectively transform as $(\mathbf{2}, 1/2, \underline{3}, i, -1, \om^{4})\equiv (H\phi_l\chi)_{\underline{3}}$ and $(\mathbf{2}, 1/2, \underline{3}, i, 1, \om^{3})$ $\equiv (H\phi_l)_{\underline{3}}\sim (H^'\phi_l \rho)_{\underline{3}}$ are required to construct invariant terms which generate the charged-lepton mass matrix. The left-handed Majorana mass terms can be created by the couplings of $\overline{\psi}_{L} \psi^c_{L}\sim (\mathbf{1}, 1, \underline{1}+ \underline{1}^'+\underline{1}^{''}+\underline{3}_s+\underline{3}_a, 1, 1, \om^{2})$ to scalars which are all prevented by $Z_4$ and $Z_8$ symmetries which are listed in Appendix \ref{forbidappen}.
The Dirac neutrino mass terms can be produced through the couplings of $\overline{\psi}_{L} \nu_{1,2,3R}$ to scalars in which $\overline{\psi}_L \nu_{1,3R}\sim (\mathbf{2}, 1/2, \underline{3}, -1, -1, \om)$ and $\overline{\psi}_L \nu_{2R}\sim (\mathbf{2}, 1/2, \underline{3}, 1, -1, \om)$. Therefore, the scalar doublets which respectively transform as $(\mathbf{2}, -1/2, \underline{3}, -1, -1, \om^{7})\equiv (\widetilde{H}\phi^*_\nu)_{\underline{3}}\sim (\widetilde{H^'}\phi^*_\nu \rho^*)_{\underline{3}}$ and $(\mathbf{2}, -1/2, \underline{3}, 1, -1, \om^{7})$ $\equiv (\widetilde{H}\phi_\nu)_{\underline{3}}\sim (\widetilde{H^'}\phi_\nu \rho^*)_{\underline{3}}$ are needed to form invariant terms which generate the Dirac neutrino mass matrix.
The 
mass terms formed by the couplings of $\overline{\psi}_{L} \nu_s\sim (\mathbf{2}, 1/2, \underline{3}, 1, 1, \om^{4})$ to scalars are all forbidden by one of the additional symmetries $A_4, Z_4, Z_2$ and $Z_8$ which are listed in Appendix \ref{forbidappen}.
The right-handed Majorana neutrino mass terms can be produced by the couplings of $\nu^c_{iR} \nu_{jR}\, (i,j=1\div 3)$ to scalars in which $\nu^c_{kR} \nu_{kR}\, (k=1\div3), \nu^c_{1R} \nu_{3R}$ and $\nu^c_{3R} \nu_{1R}$ by themselves are invariant under all the considered symmetries while $\overline{\nu}^C_{1R} \nu_{2R}\sim \overline{\nu}^C_{2R} \nu_{1R}\sim\overline{\nu}^C_{2R} \nu_{3R}\sim\overline{\nu}^C_{3R} \nu_{2R}\sim (\mathbf{1}, 0, \underline{1}, -1, 1, 1)$ which can couple to  $\big(\phi^2_\nu\big)_{\underline{1}}\sim \big(\phi^{*2}_\nu\big)_{\underline{1}}\sim (\textbf{1}, 0, \underline{1}, -1, 1, 1)$ to form invariant terms that generate the Majorana neutrino mass matrix. The sterile neutrino mass terms can be generated via the couplings of $\overline{\nu}^c_{s} \nu_{1,2,3R}$ to scalars in which the mass terms formed by the couplings of $\overline{\nu}^c_{s} \nu_{1,3R}\sim (\mathbf{1}, 0, \underline{1}, -1, -1, \om^3)$ to scalars are prevented by one of the additional symmetries $A_4, Z_4$ and $Z_8$ which are listed in Appendix \ref{forbidappen} while 
$\overline{\nu}^c_{s} \nu_{2R}\sim (\mathbf{1}, 0, \underline{1}, 1, -1, \om^3)$ can couple to 
$\left(\chi^{*3}\right)_{\underline{1}}\sim (\mathbf{1}, 0, \underline{1}, 1, -1, \om^5)$ to produce the sterile neutrino mass matrix. Finally, the mass terms formed by the couplings of $\overline{\nu}^c_{s} \nu_s\sim (\mathbf{1}, 0,  \mathbf{1}, 1, 1, \om^6)$ to scalars are all forbidden by one of the additional symmetries $A_4, Z_4$ and $Z_8$ which are listed in Appendix \ref{forbidappen}.
\begin{table}[ht]
\centering
\caption{\label{particlecont}Particle content of the model. Here $\psi_L\equiv \psi_{1,2,3L}$ and $\om=e^{i\frac{\pi}{4}}$.}
\vspace{0.25 cm}
\begin{tabular}{|c|c|c|c|c|c|c|c|c|c|c|c|c|c|c|c|c|c|c|c|c|c|}
\hline
& $\psi_L$ &$l_{1R}$&$l_{2,3R}$&$H, H^{\prime}$&$\phi_l, \phi_\nu$ &$\chi, \rho$& $\nu_{1,3R}$&$\nu_{2R}$&$\nu_s$ \\\hline
$\big[SU(2)_L, \mathrm{U}(1)_Y\big]$ & $\big[2, -\frac{1}{2}\big]$&$\big[1,-1\big]$& $\big[1,-1\big]$&$\big[2,\frac{1}{2}]$&
 $[1,0]$ &$[1,0]$&$[1,0]$&$[1,0]$& $[1,0]$   \\
$A_4$& $\underline{3}$ &$\underline{1}$&$\underline{1}^{''}$ &$\underline{1}^{''}, \underline{1}$ &
$\underline{3}$ &$\underline{1}^', \underline{1}$ &$\underline{1}$ &$\underline{1}$&$\underline{1}$ \\ 
$Z_4$&$1$ &$-i$&$-i$ &$i, 1$ &$1, i$ &$1, i$ &$-1$ &$1$&$1$  \\
$Z_2$&$-1$ &$1$&$-1$&$-1$&$-1, 1$ &$-1, 1$ &$1$ &$1$& $-1$\\
$Z_8$&$\om^7$&$\om^3$&$\om^4$&$\om, \om^3$&$\om^2, 1$&$\om, \om^6$&$1$&$1$&$\om^3$ \\\hline
\end{tabular}
\end{table}

The particle content in Table \ref{particlecont} yields the following up to 6D Yukawa interactions invariant
under all the model symmetries:
\begin{equation}
-\mathcal{L} = -\mathcal{L}^{(L)} -\mathcal{L}^{(D)} - \mathcal{L}^{(R)} - \mathcal{L}^{(S)} + h.c.,
\end{equation}
where\footnote{It is noted that $\left(\overline{\nu}^c_s\nu_{2R}\right)_{\underline{1}}\left(\phi^2_l\chi\right)_{\underline{1}}$ and $\left(\overline{\nu}^c_s\nu_{2R}\right)_{\underline{1}}\left(\phi^{*2}_l\chi\right)_{\underline{1}}$ are invariant under all the considered symmetries, however, with the VEV alignment of $\phi_l$ in Eq. (\ref{VEV}), $\left(\phi^{2}_l\right)_{\underline{1}^{''}}=\left(\phi^{*2}_l\right)_{\underline{1}^{''}}=0$; thus, $\left(\phi^{2}_l \chi\right)_{\underline{1}}=\left(\phi^{*2}_l\chi\right)_{\underline{1}}=0$ as a consequence of the tensor product $\underline{3}\times \underline{3}$ of $A_4$ in the T-diagonal basis.},
\bea
-\mathcal{L}^{(L)} &=& \frac{h_1}{\Lambda^2}(\overline{\psi}_L \phi_l)_{\underline{1}} \left(H\chi l_{1R}\right)_{\underline{1}}
+\frac{h_2}{\Lambda}(\overline{\psi}_L \phi_l)_{\underline{1}^{''}} (H l_{2R})_{\underline{1}^{'}}
+ \frac{h_3}{\Lambda}(\overline{\psi}_L \phi_l)_{\underline{1}^{''}} (H l_{3R})_{\underline{1}^{'}}\crn
&+&\frac{h_4}{\Lambda^2}\Big[(\overline{\psi}_L \phi_l)_{\underline{1}^{'}} (H^' \rho l_{2R})_{\underline{1}^{''}}
+ (\overline{\psi}_L \phi_l)_{\underline{1}^{'}} (H^' \rho l_{3R})_{\underline{1}^{''}}\Big]+ h.c, \label{Lyclep}\\
-\mathcal{L}^{(D)} &=&\frac{x_1}{\Lambda}(\overline{\psi}_L \phi^*_\nu)_{\underline{1}^{''}} (\widetilde{H}  \nu_{1R})_{\underline{1}^{'}} + \frac{x_2}{\Lambda}(\overline{\psi}_L \phi_\nu)_{\underline{1}^{''}} (\widetilde{H} \nu_{2R})_{\underline{1}^{'}}
+ \frac{x_3}{\Lambda}(\overline{\psi}_L \phi^*_\nu)_{\underline{1}^{''}} (\widetilde{H}\nu_{3R})_{\underline{1}^{'}}\crn
&+&\frac{x_4}{\Lambda^2}(\overline{\psi}_L \phi^*_\nu)_{\underline{1}} (\widetilde{H}^' \rho^* \nu_{1R})_{\underline{1}}
+ \frac{x_5}{\Lambda^2}(\overline{\psi}_L \phi_\nu)_{\underline{1}} (\widetilde{H}^' \rho^* \nu_{2R})_{\underline{1}} +\frac{x_6}{\Lambda^2}(\overline{\psi}_L \phi^*_\nu)_{\underline{1}} (\widetilde{H}^' \rho^* \nu_{3R})_{\underline{1}}+h.c,  \label{LD}\eea
\bea
-\mathcal{L}^{(R)} &=&\frac{\lambda_1}{2} \left(\overline{\nu}_{1R}^c \nu_{1R}+\overline{\nu}_{2R}^c \nu_{2R}+\overline{\nu}_{3R}^c \nu_{3R}\right)_{\underline{1}} +\frac{\lambda_2}{2} \left(\overline{\nu}_{1R}^c\nu_{3R}+\overline{\nu}_{3R}^c\nu_{1R}\right)_{\underline{1}} \crn
&+& \frac{y}{2\Lambda}\left(\overline{\nu}_{1R}^c\nu_{2R}+\overline{\nu}_{2R}^c\nu_{1R}+\overline{\nu}_{2R}^c\nu_{3R}+\overline{\nu}_{3R}^c\nu_{2R}\right)_{\underline{1}} \left(\phi^2_\nu+\phi^{*2}_\nu\right)_{\underline{1}} +h.c, \label{LR}\\
-\mathcal{L}^{(S)}&=&\frac{z}{\Lambda^2}\left(\overline{\nu}^c_s\nu_{2R}\right)_{\underline{1}}\left(\chi^{*3}\right)_{\underline{1}} + h.c, \label{LS}
\eea
with $\widetilde{H} = i \sigma_2 H^*$, $\Lambda$ being the cut-off scale, $h_{i}, y_j\, (i=1\div 4; j=1\div 6)$ and $z$ are the Yukawa-like couplings, and $\lambda_{1, 2}$ are the Majorana mass scales. Each of additional Abelian symmetry $Z_2, Z_4$ and $Z_8$ takes a crucial role in preventing the unwanted mass terms in the Lagrangian to obtain the desired lepton mass matrices which are listed Appendix \ref{forbidappen}. As an example, in the absence of $Z_4\rtimes Z_4$, there will be additional invariant terms $(\overline{\psi}_{1L} l_{1R})_{1_{+0}} H_2$ and $(\overline{\psi}_{1L} l_{1R})_{1_{+0}} H_3$ which contribute to the entry "11" of the charged lepton matrix. If Yukawa-like couplings corresponding to $(\overline{\psi}_{1L} l_{1R})_{1_{+0}} H_2$ and $(\overline{\psi}_{1L} l_{1R})_{1_{+0}} H_3$ are of the same order of magnitude as that of $x_1$, the contribution of these terms is about $10^3$ times larger than the contribution\footnote{In this study, $\Lambda=10^{13} \, \mathrm{GeV},\, v_\chi = 10^{10} \, \mathrm{GeV}$, i.e., $v_\chi/\Lambda \sim 10^{-3}$.} of $\frac{1}{\Lambda}\left(\bar{\psi}_{1L}l_{1 R}\right)_{1_{-0}} (H_1\chi)_{1_{-1}}$. As a consequence, the charged-lepton mass hierarchy cannot naturally account for the experimental data \cite{PDG2022}.

We consider the following vacuum expectation value (VEV) alignments for scalar fields,
\bea
&&\langle H\rangle =\left(0\hs\hs v\right)^T, \hs\hs \langle H^'\rangle =\left(0\hs\hs v^'\right)^T,  \crn
&&\langle\phi_l\rangle =(v_l,\,\, 0,\,\, 0),\hs   \langle\phi_\nu\rangle=(v_{1}, \,\, v_{2}, \,\, v_{3}), \hs \langle\chi\rangle =v_\chi, \hs  \langle\rho\rangle = v_\rho.\label{VEV}
\eea
The Higgs doublet VEVs $v$ and $v^'$ obey the total electroweak VEV, $v^2+v^{'2}=v^2_{\mathrm{w}}=(246\, \mathrm{GeV})^2$; thus, they can be conveniently parametrized as\footnote{For shortly, we employ the following notations: $s_\beta=\sin \beta,\, c_\beta=\cos\beta, \, s_\psi=\sin \psi,\, c_\psi=\cos \psi, \, s_\alpha=\sin\alpha, 
\, s_\theta=\sin \theta, \, s_\delta=\sin\delta_{CP}, \, s^'_\delta=\sin\delta^'_{CP}, \, s_{ij}=\sin\theta_{ij},\, c_{ij}=\cos\theta_{ij}, \, s^'_{ij}=\sin\theta^'_{ij}$ and $c^'_{ij}=\cos\theta^'_{ij} \, (ij=12,23,13)$.}
\bea
&&v=v_{\mathrm{w}} s_\beta,\hs   v^'=v_{\mathrm{w}} c_\beta. \label{higgvevs}\eea
Furthermore, in order to have heavy right handed Majorana neutrino masses, and hence allowing the implementation of the type I seesaw mechanism that generates the small masses of the active neutrinos, the VEVs of flavons $\phi_l,\phi_\nu,\rho, \chi$ and the cut-off scale $\Lambda$ are
at a very high scale:
\bea
&&v_\chi\simeq 5\times 10^{9} \, \mathrm{GeV}, \hs  v_\rho \sim v_l \simeq  5\times 10^{11} \, \mathrm{GeV}, \hs \Lambda\simeq 10^{13} \, \mathrm{GeV}, \crn
&&v_{1\nu} \simeq 10^{11}\, \mathrm{GeV}, \hs v_{2\nu} \simeq 2\times 10^{11}\, \mathrm{GeV}, \hs v_{3\nu} \simeq 3\times 10^{11}\, \mathrm{GeV}. \label{flavonscale}
\eea
In models with more than one $SU(2)_L$ doublet as in this study, the FCNC processes such as $b\rightarrow s \gamma$ exist in the Higgs sector. However, they are suppressed by non-Abelian discrete symmetries \cite{Mondragon07,Kubo13}. To make such processes below the experimental bounds, some restrictions on the model parameters such as the large masses for non SM scalars and Yukawa couplings need to be imposed. This kind of the model contains many free parameters which allows us freedom to assume that the remaining scalars are sufficiently heavy to fullfil the current experimental bounds.
Furthermore, the off-diagonal Yukawa couplings in the charged-lepton sector, Eq. (\ref{Lyclep}), are proportional to $\frac{v_{l}}{\La}\sim 10^{-2}$. Therefore, the LFV processes, such as $l_j\rightarrow l_i \gamma$, are suppressed by the tiny factor $\fr{v_{l}}{\La}\fr{1}{m_H^2}$ associated
with the small Yukawa couplings and the large mass scale of the heavy scalars $m_H$ \cite{Dorsner15, Davidson10, Davidson16,Vien2021}.
\section{\label{neutrinomixing} Neutrino mass and mixing}
\subsection{Lepton mass and mixing in the three neutrino scheme}
From the Yukawa terms in Eq. (\ref{Lyclep}), by using the multiplication rules of the $A_4$ group in the $T$-diagonal basis \cite{ishi}, when the scalar fields $H, H^', \phi_l, \chi$ and $\rho$ get the VEVs in Eqs. (\ref{VEV}) and (\ref{higgvevs}), we obtain the following charged lepton mass matrix
\bea
M_L &= \left(\begin{matrix}
a_{l} & 0 & 0 \\
0 & d_l & d_l \\
0 & b_l & c_l
\end{matrix}\right),
\eea
with
\bea
&&a_l=h_1 v_{\mathrm{w}} s_\beta \frac{v_\chi}{\Lambda} \frac{v_l}{\Lambda}, \hs
b_l=h_2 v_{\mathrm{w}} s_\beta \left(\frac{v_l}{\Lambda}\right), \hs
c_l=h_3 v_{\mathrm{w}} s_\beta \left(\frac{v_l}{\Lambda}\right), \hs
d_l= h_4 v_{\mathrm{w}} c_\beta  \frac{v_l}{\Lambda} \frac{v_\rho}{\Lambda}. \label{abcdl}
\eea
In general, $a_l, b_l, c_l$ and $d_l$ are complex parameters, thus, $M_L$ is a complex matrix. For simplicity, we consider the case of $\arg c_l =\arg b_l$. The Hermitian matrix $m_l=M_L M^+_L$, whose real and positive eigenvalues, given by
\bea
m_l&=&\left(
\begin{array}{ccc}
 a_0^2 & 0 & 0 \\
 0 & 2d_0^2 & \big(b_{0}+c_{0}\big) d_{0}e^{-i \al} \\
 0 & \big(b_{0}+c_{0}\big) d_{0} e^{i \al} & b_{0}^2+c_{0}^2 \\
\end{array}
\right), \label{ml}
\eea
where $\al=\arg b_l-\arg d_l$ while $a_0, b_0, c_0, d_0$ and $\al_a, \al_b, \al_c, \al_d$ are\footnote{$\al_a$ is absorbed during the construction of the matrix $m_l$.} respectively the magnitudes and the arguments of $a_l, b_l, c_l, d_l$.
The matrix $m_l$ is diagonalised by the matrices $U_{l, r}$ satisfying
\bea
U^\dag_{l} m_l U_{r}=\mathrm{diag} (m^2_e, m^2_\mu, m^2_\tau), \eea
where
\bea
&&m^2_e =a^2_0, \hs
m^2_{\mu, \tau}= \frac{1}{2} \left(b_{0}^2+c_{0}^2+2d_{0}^2\mp \sqrt{(b_0^2 + c_0^2)^2 + 4 (2 b_0 c_0 + d_0^2) d_0^2}\right),  \label{memt}\\
&&U_{l}=U_{r}=\left(%
\begin{array}{ccc}
  1 & 0 & 0 \\
   0 & c_\psi &\,\,\,\,\,\, s_\psi\, e^{-i \al}\\
  0 & -s_\psi\, e^{i \al} & c_\psi  \\
\end{array}%
\right), \label{Uclep} \\
&&s_\psi=\frac{1}{\sqrt{1+\frac{\left(b_0^2+c_{0}^2-2d_{0}^2+\sqrt{(b_0^2 + c_0^2)^2 + 4 (2 b_0 c_0 + d_0^2) d_0^2}\right)^2}{4 (b_{0}+c_{0})^2 d^2_0}}}\hs (c_\psi>0).\label{psi}
\eea
In the charged-lepton sector, there exist six parameters including $h_{1,2,3,4}, s_\beta$ and $c_\alpha$. Since $c_\alpha$ is determined in neutrino sector, Eq. (\ref{UL}); thus, leaving five parameters $h_{1,2,3,4}$ and $s_\beta$ corresponding to the three observed experimental parameters $m_e, m_\mu$ and $m_\tau$.

Expressions (\ref{abcdl}) and (\ref{memt}) yields the following relations:
\bea
&&h_{1}=\frac{\La^2 m_{e}}{ v _{\chi} v_l v_{\mathrm{w}} s_\beta}, \hs h_{2,3}=\frac{\sqrt{\delta_0}\mp s_\beta c_\beta h_4 \La^5 m_\tau m_\mu  v_{\mathrm{w}}^2 v_l^2 v_\rho}{2 s_\beta^2  c_\beta^2 h_4^2 \La^2 v_{\mathrm{w}}^4 v_l^4 v_\rho^2}, \label{h1h2h3h4relation}
\eea
where
\bea
\delta_0=-s_\beta^2  c_\beta^2 h_4^2 \La^2 v_{\mathrm{w}}^4 v_l^4 v_\rho^2 (\La^4 m_\tau^2 -
   2 c_\beta^2 h_4^2 v_{\mathrm{w}}^2 v_l^2 v_\rho^2) (\La^4 m_\mu^2 -
   2 c_\beta^2 h_4^2 v_{\mathrm{w}}^2 v_l^2 v_\rho^2). \label{del0}
\eea
Equation (\ref{Uclep}) tells us that $U_{l}$ is non trivial
and hence it will affect on
the lepton mixing matrix.

Next, we consider the neutrino sector. When the scalars get the VEVs in Eqs. (\ref{VEV}) and (\ref{higgvevs}), we get the following Dirac, Majorana and sterile mass matrices:
\bea
&&M_D = \left(\begin{matrix}
a_{31} & b_{31} & c_{31} \\
a_{23} &b_{23} & c_{23} \\
a_{12} & b_{12} & c_{12}
\end{matrix}\right), \hs\hs M_R = \left(\begin{matrix}
\lambda_1 & d & \lambda_2 \\
d & \lambda_1 & d \\
\lambda_2 & d & \lambda_1
\end{matrix}\right), \hs\hs M_S= \left(0\hs \hs g \hs \hs 0\right), \label{MDRS}
\eea
where
\bea
&&a_{12}=\frac{x_1}{\Lambda} v v_{1\nu} +\frac{x_4}{\Lambda^2} v^' v_\rho v_{2\nu}, \, a_{23}=\frac{x_1}{\Lambda} v v_{2\nu}+\frac{x_4}{\Lambda^2} v^' v_\rho v_{3\nu},\, a_{31}=\frac{x_1}{\Lambda} v v_{3\nu}+\frac{x_4}{\Lambda^2} v^' v_\rho v_{1\nu}, \hs\hs \crn
&&b_{12}=\frac{x_2}{\Lambda} v v_{1\nu} +\frac{x_5}{\Lambda^2} v^' v_\rho v_{2\nu},\, b_{23}=\frac{x_2}{\Lambda} v v_{2\nu}+\frac{x_5}{\Lambda^2} v^' v_\rho v_{3\nu},\,b_{31}=\frac{x_2}{\Lambda} v v_{3\nu}+\frac{x_5}{\Lambda^2} v^' v_\rho v_{1\nu}, \hs\hs \crn 
&&c_{12}=\frac{x_3}{\Lambda} v  v_{1\nu} +\frac{x_6}{\Lambda^2} v^' v_\rho v_{2\nu},\, c_{23}=\frac{x_3}{\Lambda} v v_{2\nu} +\frac{x_6}{\Lambda^2} v^' v_\rho v_{3\nu},\,c_{31}=\frac{x_3}{\Lambda} v v_{3\nu} +\frac{x_6}{\Lambda^2} v^' v_\rho v_{1\nu}, \hs\hs \label{aDij}\\
&&d=\frac{2y}{\Lambda} \left(v^2_{1\nu}+2v_{2\nu}v_{3\nu}\right), \hs g=\frac{z v^3_\chi}{\Lambda^2}.\label{dg}
\eea
In the $(\nu^c_{L},\nu_{R}, \nu_s)$ basis, the $(7\times 7)$ neutrino mass matrix takes the form:
\begin{equation}
M_{\nu}^{7\times 7} = \left( \begin{matrix}
0 & M_D & 0 \\
M_D^T & M_R & M_S^T \\
0 & M_S & 0
\end{matrix} \right). \label{mnu77}
\end{equation}
Expressions (\ref{higgvevs}), (\ref{flavonscale}), (\ref{aDij}) and (\ref{dg}) inly that\footnote{With the help of Eq. (\ref{l1l2}).}  the Dirac and sterile neutrino masses are much smaller than the right-handed neutrino ($M_D < M_S \ll M_R $); thus, one can block-diagonalise the $7\times 7$ matrix by using the seesaw formula and get the effective $4\times 4$ light neutrino mass matrix in the $(\nu^c_L, \nu_s)$ basis \cite{Barry2011v}
\begin{equation}
M_{\nu}^{4\times 4} = -\left(\begin{matrix}
M_DM_R^{-1}M_D^T & M_DM_R^{-1}M_S^T \\

M_S(M_R^{-1})^TM_D^T & M_SM_R^{-1}M_S^T
\end{matrix} \right).
\label{mnu44}
\end{equation}
By applying the type-I seesaw mechanism, the  $3\times 3$ active
neutrino mass matrix is obtained as\cite{Barry2011v}
\begin{equation}
m_{\nu} \simeq M_DM_R^{-1}M_S^T\left(M_S M_R^{-1}M_S^T\right)^{-1}M_S\left(M_R^{-1}\right)^T M_D^T-M_DM_R^{-1}M_D^T.\label{mnu33}
\end{equation}
Substituting Eq. (\ref{MDRS}) into Eq. (\ref{mnu33}) yields:
\bea
 m_\nu &=&\left(
\begin{array}{ccc}
 A_1 & A_{12} & A_{13} \\
 A_{12} & A_{2} & A_{23} \\
 A_{13} & A_{23} & A_{3} \\
\end{array}
\right),\label{mnu33v}
\eea
where
\bea
&&A_{1}=\frac{(a_{31}^2+c_{31}^2) \lambda_{1}-2 a_{31} c_{31} \lambda_{2}}{\lambda_{2}^2-\lambda_{1}^2}, \hs
A_{12}=\frac{(a_{23} a_{31}+c_{23} c_{31})\lambda_{1}-(a_{23} c_{31} +a_{31} c_{23}) \lambda_{2}}{\lambda_{2}^2-\lambda_{1}^2},\crn
&&A_{2}=\frac{(a_{23}^2+c_{23}^2) \lambda_{1}-2 a_{23} c_{23} \lambda_{2}}{\lambda_{2}^2-\lambda_{1}^2}, \hs
A_{23}=\frac{(a_{12} a_{23}+c_{12} c_{23}) \lambda_{1}-(a_{12} c_{23}+a_{23} c_{12})\lambda_{2}}{\lambda_{2}^2-\lambda_{1}^2}, \crn
&&A_{3}=\frac{(a_{12}^2+c_{12}^2)\lambda_{1}-2 a_{12} c_{12} \lambda_{2}}{\lambda_{2}^2-\lambda_{1}^2},  \hs
A_{13}=\frac{(a_{12} a_{31} +c_{12} c_{31})\lambda_{1}-(a_{12} c_{31}+a_{31} c_{12})\lambda_{2}}{\lambda_{2}^2-\lambda_{1}^2}. \label{Aij}
\eea
The expressions (\ref{aDij}), (\ref{dg}) and (\ref{Aij}) show that, in general the $3\times 3$ active neutrino mass matrix $m_\nu$ in Eq. (\ref{mnu33}) is a complex matrix. Considering the case of real VEVs for the scalar fields $H, H^', \phi_l, \phi_\nu, \chi$ and $\rho$, and the phase redefinition of lepton fields allows to rotate away the phases of three Yukawa couplings $x_{1\div 6}$, $y$ and $z$  that make the mass matrix $m_\nu$ real and can be diagonalized by the unitary matrix $U_{\nu}$. 
The mass matrix $m_{\nu}$ in Eq.(\ref{mnu33v}) has three eigenvalues and corresponding eigenvectors as follows
\bea
&&m_1 =0,\hs m_{2,3}=\kappa_1 \mp \kappa_2,\label{m123}\\
&&\varphi_{l}=\left( \frac{k_{l}}{\sqrt{k_{l}^2+n_{l}^2+1}},
 \frac{n_l}{\sqrt{k_l^2+n_l^2+1}},
 \frac{1}{\sqrt{k_l^2+n_l^2+1}}
\right)^T,\label{eigenvectors}
\eea
where $l=1,2,3$ and the explicit expressions of $\kappa_{1, 2}$, $k_{1,2,3}$ and $n_{1,2,3}$ are given in Appendix \ref{Appkappakn12}, which obey the following relations
 \bea
 && k_1 k_2 + n_1 n_2+1=0,\,\,  k_1 k_3 + n_1 n_3+1=0,\,\, k_2 k_3 + n_2 n_3+1=0.  \label{kn12relat}
 \eea
It is noted that the eigenvalue $m_1=0$ corresponds to the first neutrino eigenvector $\varphi_1$. Thus, the neutrino mass should be either $\left(0,\, m_2,\, m_3\right)$ or $\left(m_2,\, m_3,\, 0\right)$. The considered model can accommodate both inverted ordering (IO) and normal ordering (NO) being consistent with the experimental data and different from that of Refs. \cite{Krishnan2020,Singh2022} whereby only the NO is allowed. The eigenvalues and eigenvectors of $ m_{\nu}$ in Eq. (\ref{mnu33v}) corresponding two mass hierarchies are given by\footnote{At present, there is no solid evidence that
all neutrinos have non-zero mass, and one of them may be exactly massless.}:
\bea
U_{\nu }^T m_{\nu} U_{\nu }=\left\{
\begin{array}{l}
\left(%
\begin{array}{ccc}
0\quad & 0 &0 \\
0\quad  & m_2 & 0 \\
0\quad & 0 & m_3\\
\end{array}%
\right),\hspace{0.1cm} U_{\nu }=\left(
\begin{array}{ccc}
 \frac{k_{1}}{\sqrt{k_{1}^2+n_{1}^2+1}} & \frac{k_{2}}{\sqrt{k_{2}^2+n_{2}^2+1}} & \frac{k_{3}}{\sqrt{k_{3}^2+n_{3}^2+1}} \\
 \frac{n_{1}}{\sqrt{k_{1}^2+n_{1}^2+1}} & \frac{n_{2}}{\sqrt{k_{2}^2+n_{2}^2+1}} & \frac{n_{3}}{\sqrt{k_{3}^2+n_{3}^2+1}} \\
 \frac{1}{\sqrt{k_{1}^2+n_{1}^2+1}} & \frac{1}{\sqrt{k_{2}^2+n_{2}^2+1}} & \frac{1}{\sqrt{k_{3}^2+n_{3}^2+1}} \\
\end{array}
\right) \hspace{0.2cm}\mbox{for NO,}\ \  \\
\left(%
\begin{array}{ccc}
m_2& 0 &\hspace{0.1 cm} 0 \\
0 & m_3  &\hspace{0.1 cm}  0 \\
0& 0 & \hspace{0.1 cm} 0\\
\end{array}%
\right),\hspace{0.1cm} U_{\nu }=\left(
\begin{array}{ccc}
\frac{k_{2}}{\sqrt{k_{2}^2+n_{2}^2+1}} & \frac{k_{3}}{\sqrt{k_{3}^2+n_{3}^2+1}}&\frac{k_{1}}{\sqrt{k_{1}^2+n_{1}^2+1}} \\
\frac{n_{2}}{\sqrt{k_{2}^2+n_{2}^2+1}} & \frac{n_{3}}{\sqrt{k_{3}^2+n_{3}^2+1}}& \frac{n_{1}}{\sqrt{k_{1}^2+n_{1}^2+1}}  \\
\frac{1}{\sqrt{k_{2}^2+n_{2}^2+1}} & \frac{1}{\sqrt{k_{3}^2+n_{3}^2+1}} & \frac{1}{\sqrt{k_{1}^2+n_{1}^2+1}} \\
\end{array}
\right) \hspace{0.2cm}\mbox{for IO.}
\end{array}%
\right.  \label{Unu}
\eea
From Eqs. (\ref{m123}) and (\ref{Unu}), we can express the model parameters $\kappa_1$ and $\kappa_2$ in terms of two observed parameters $\Delta m^2_{21}$ and $\Delta m^2_{31}$ as follows:
\bea
&&\kappa_1=\left\{
\begin{array}{l}
\frac{1}{2}\left(\sqrt{\Delta m^2_{31}}+\sqrt{\Delta m^2_{21}}\right)\hspace{0.2cm}\mbox{for\, NO},    \\
\frac{1}{2} \sqrt{2 \sqrt{\Delta m^2_{31} (\Delta m^2_{31}-\Delta m^2_{21})}+\Delta m^2_{21}-2 \Delta m^2_{31}} \hspace{0.1cm}\,\mbox{for\,  IO},
\end{array}%
\right. \label{k1}\\
&&\kappa_{2}=\left\{
\begin{array}{l}
\frac{1}{2}\left(\sqrt{\Delta m^2_{31}}-\sqrt{\Delta m^2_{21}}\right)\hspace{0.3cm}\mbox{for \, NO}.    \\
\frac{\Delta m^2_{21}}{2 \sqrt{2 \sqrt{\Delta m^2_{31} (\Delta m^2_{31}-\Delta m^2_{21})}+\Delta m^2_{21}-2 \Delta m^2_{31}}} \hspace{0.1cm}\,\mbox{for \, IO}.
\end{array}
\right. \label{k2}\eea
The $3\times 3$ leptonic mixing matrix, $U_{L}=U_{l}^{\dag} U_{\nu }$, is
\bea
U_{L}=\left\{
\begin{array}{l}
\left(
\begin{array}{ccc}
 \frac{k_{1}}{\sqrt{k_{1}^2+n_{1}^2+1}} & \frac{k_{2}}{\sqrt{k_{2}^2+n_{2}^2+1}} & \frac{k_{3}}{\sqrt{k_{3}^2+n_{3}^2+1}} \\
 \frac{n_{1} \cos\psi-e^{-i \al}  \sin\psi}{\sqrt{k_{1}^2+n_{1}^2+1}} & \frac{n_{2} \cos\psi-e^{-i \al} \sin\psi}{\sqrt{k_{2}^2+n_{2}^2+1}} & \frac{n_{3} \cos\psi-e^{-i \al}  \sin\psi}{\sqrt{k_{3}^2+n_{3}^2+1}} \\
 \frac{ \cos\psi+e^{i \al} n_{1} \sin\psi}{\sqrt{k_{1}^2+n_{1}^2+1}} & \frac{ \cos\psi+e^{i \al} n_{2} \sin\psi}{\sqrt{k_{2}^2+n_{2}^2+1}} & \frac{ \cos\psi+e^{i \al} n_{3} \sin\psi}{\sqrt{k_{3}^2+n_{3}^2+1}} \\
\end{array}
\right) \hspace{0.1cm}\mbox{for  NO},  \label{UL}  \\
\left(
\begin{array}{ccc}
 \frac{k_{2}}{\sqrt{k_{2}^2+n_{2}^2+1}} & \frac{k_{3}}{\sqrt{k_{3}^2+n_{3}^2+1}}&\frac{k_{1}}{\sqrt{k_{1}^2+n_{1}^2+1}}  \\
 \frac{n_{2} \cos\psi-e^{-i \al} \sin\psi}{\sqrt{k_{2}^2+n_{2}^2+1}} & \frac{n_{3} \cos\psi-e^{-i \al}  \sin\psi}{\sqrt{k_{3}^2+n_{3}^2+1}} &\frac{n_{1} \cos\psi-e^{-i \al}  \sin\psi}{\sqrt{k_{1}^2+n_{1}^2+1}} \\
 \frac{ \cos\psi+e^{i \al} n_{2} \sin\psi}{\sqrt{k_{2}^2+n_{2}^2+1}} & \frac{ \cos\psi+e^{i \al} n_{3} \sin\psi}{\sqrt{k_{3}^2+n_{3}^2+1}} &\frac{ \cos\psi+e^{i \al} n_{1} \sin\psi}{\sqrt{k_{1}^2+n_{1}^2+1}} \\
\end{array}
\right) \hspace{0.1cm}\mbox{for IO}.
\end{array}%
\right.
\eea
Comparing Eq. (\ref{UL}) with the standard parameterization of the lepton mixing matrix $\mathrm{U_{PMNS}}$, with the help of Eq. (\ref{kn12relat}), yields the relations between the model parameters and three observed mixing angles\footnote{For simplicity, hereafter we will use the notations:
$c_{ij}=\cos \theta_{ij},\, s_{ij}=\sin \theta_{ij}\, (ij=12, 23, 13);\, S^2_{k4}=\sin^2 \theta_{k4}\, (k=1, 2,3)$, and $c_\alpha=\cos\alpha, \hs s_\alpha=\sin\alpha, \hs c_\psi=\cos\psi, \hs s_\psi=\sin\psi$.} 
$s_{12}, s_{23}$ and $s_{13}$:
\bea &&s_{13}^2=\left| \mathrm{U}_{e 3}\right|^2=\left\{
\begin{array}{l}
\frac{k_3^2}{1 + k_3^2 + n_3^2}\hspace{0.3cm}\mbox{for \, NO},    \\
\frac{k_{1}^2}{k_{1}^2+n_{1}^2+1}\hspace{0.25cm}\,\mbox{for \, IO},
\end{array}%
\right.  \label{s13sq}\\
&& s_{12}^2 =\frac{\left| \mathrm{U}_{e 2}\right|^2}{1-\left| \mathrm{U}_{e 3}\right|^2}=
\left\{
\begin{array}{l}
\frac{(1 + n_2 n_3)^2 (1 + k_3^2 + n_3^2)}{(1 + n_3^2) \left[k_3^2 (1 + n_2^2) + (1 + n_2 n_3)^2\right]}\hspace{0.3cm}\mbox{for \, NO},    \\
\frac{\left(k_{1} k_{2}+n_{1}^2+1\right)^2}{\left(n_{1}^2+1\right) \left[(k_{1} k_{2}+1)^2+\left(k_{2}^2+1\right) n_{1}^2\right]}\hspace{0.25cm}\,\mbox{for \, IO},
\end{array}%
\right. \hspace{0.25cm}, \label{s12s13relation}\\
&& s_{23}^2=\frac{\left| \mathrm{U}_{\mu 3}\right|^2}{1-\left| \mathrm{U}_{e 3}\right|^2}=
\left\{
\begin{array}{l}
\fr{n_3^2 c^2_\psi+ s^2_\psi - n_3 s_{2\psi} c_\alpha}{n^2_3+1}\hspace{0.3cm}\mbox{for \, NO},    \\
\frac{n_{1}^2 c^2_\psi +s^2_\psi-n_{1} s_{2_\psi} c_\alpha}{n_{1}^2+1} \hspace{0.25cm}\,\mbox{for \, IO},
\end{array}%
\right. \label{s23sq}\eea
where $k_{1,2,3}$ and $n_{1,2,3}$ are defined in Appendix \ref{Appkappakn12}.
Therefore, we can express the model parameters $k_{1,2,3}$ and $n_{1,2,3}$ in terms of three observed parameters $s_{12}, s_{13}$, $s_{23}$ and two constrained parameters $\psi, \alpha$ as follows:
\begin{itemize}
  \item [$\bullet$] For NO:
\bea
&&n_{3}=
\frac{c_\psi s_\psi c_\alpha -\sqrt{s^2_{23} c^2_{23}-s^2_\alpha s^2_\psi c^2_\psi}}{c^2_\psi-s^2_{23}}, \crn
&&k_{3}=t_{13}\sqrt{\frac{c^4_\psi+s^2_\psi s^2_{23}+c^2_\psi \left(c_{2\alpha} s^2_\psi-s^2_{23}\right)- s_{2\psi} c_\alpha\sqrt{s^2_{23} c^2_{23}-s^2_\alpha s^2_\psi c^2_\psi}}{(s^2_{23}-c^2_\psi)^2}}, \crn
&&n_{2}=\frac{n_{3} \left[k_{3}^2+\left(n_{3}^2+1\right) c^2_{12}\right]-c_{12} s_{12}\left(n_{3}^2+1\right) k_{3} \sqrt{  k_{3}^2+n_{3}^2+1}}{k_{3}^2 s^2_{12}-\left(n_{3}^2+k_{3}^2+1\right) n_{3}^2 c^2_{12}}, \crn
&&k_{1}=\frac{k_{3} (n_{3}-n_{2})}{n_{2} \left(k_{3}^2+n_{3}^2\right)+n_{3}},\hs k_{2}=-\frac{n_{2} n_{3}+1}{k_{3}}, \hs n_{1}=-\frac{k_{3}^2+n_{2} n_{3}+1}{n_{2} \left(k_{3}^2+n_{3}^2\right)+n_{3}}. \label{kiniexpressionNH}
\eea
  \item [$\bullet$] For IO:
\bea
&&n_{1}=\frac{s_\psi c_\psi c_\alpha-\sqrt{s^2_{23} c^2_{23}-s^2_\alpha s^2_\psi c^2_\psi }}{c^2_\psi-s^2_{23}}, \crn
&&k_1=t_{13}\sqrt{\frac{c^4_\psi+s^2_\psi s^2_{23}+c^2_\psi \left(c_{2\alpha}s^2_\psi-s^2_{23}\right)-s_{2\psi} c_\alpha \sqrt{s^2_{23} c^2_{23}-s^2_\alpha s^2_\psi c^2_\psi}}{(s^2_{23}-c^2_\psi)^2}}, \crn
&&k_{2}=\frac{\left(n_1^2+1\right)^2 c^2_{12}}{\sqrt{\left(n_1^3+n_1\right)^2 \left(k_1^2+n_1^2+1\right)c^2_{12}s^2_{12}}- k_1 \left(n_1^2+1\right)c^2_{12}}, \crn
&&k_{3}=-\frac{k_{1} k_{2}+n_{1}^2+1}{k_{1}^2 k_{2}+k_{1}+k_{2} n_{1}^2}, \hs n_{2}=-\frac{k_{1} k_{2}+1}{n_{1}}, \hs n_{3}=\frac{n_{1} (k_{1}-k_{2})}{k_{1}^2 k_{2}+k_{1}+k_{2} n_{1}^2}. \label{kiniexpressionIH}
\eea
\end{itemize}
The Jarlskog invariant parameter in the active sector \cite{Jarlskog1}, $J_{CP}=\mathrm{Im}\left[U_{12} U_{23} U^*_{13} U^*_{22}\right]$, is given by:
\bea
J_{CP}=\frac{ k_{2} k_{3} (n_{2}-n_{3}) s_\psi c_\psi s_\alpha }{\left(k_{2}^2+n_{2}^2+1\right) \left(k_{3}^2+n_{3}^2+1\right)} \hspace{0.2 cm} \mbox{for\, both \,NO\, and\, IO.}\label{Jcp}
\eea
Combining expression (\ref{Jcp}) with that of the standard parameterization in the three neutrino scenario, $J=-c_{12} c_{13}^2 c_{23} s_{12} s_{13} s_{23} \sin \delta_{CP}$, we obtain:
\bea
\sin \delta_{CP}=\frac{k_{2} k_{3} (n_{2}-n_{3}) s_\psi c_\psi s_\alpha }{\left(k_{2}^2+n_{2}^2+1\right) \left(k_{3}^2+n_{3}^2+1\right) c_{12} c_{13}^2 c_{23} s_{12} s_{13} s_{23} } \hspace{0.15 cm} \mbox{for both NO and IO.} \label{sdv}
\eea
Two Majorana CP phases are obtained as\footnote{The diagonal matrix containing the Majorana phases has a general form of $P=\mathrm{diag}\left(e^{i\alpha_{1}}, e^{i\alpha_{2}}, e^{i\alpha_{3}}\right)$ where $\alpha_{1,2,3}$ are three Majorana phases which is specified by two combinations made of $\alpha_{1,2,3}$ such as $\alpha_{i}-\alpha_{1}$, $\alpha_{i}-\alpha_{2}$ and $\alpha_{i}-\alpha_{3}$.
We consider the case where $\alpha_i$ is specified by two combinations $\alpha_{i}-\alpha_{2}$, $P$ is reduced to $P=\mathrm{diag}\left(e^{i\gamma_1}, 1, e^{i\gamma_2}\right)$.}
\bea
&&\gamma_1=0,\hs \gamma_2=\delta_{CP} \hspace{0.2 cm} \mbox{for both NO and IO}.  \label{Majoranaphases}
\eea
The effective neutrino masses in three neutrino scheme, $m^{(3)}_{\beta} = \left(\sum^3_{i=1} \left|U_{ei}\right|^2 m_i^2 \right)^{\frac{1}{2}}$ and $\langle m^{(3)}_{ee}\rangle = \left| \sum^3_{i=1} U_{ei}^2 m_i \right|$, get the following forms \cite{betdecay2},
\bea
&&m^{(3)}_{\beta} = \left\{
\begin{array}{l}
\left(\frac{(n_{2} n_{3}+1)^2 \Delta m^2_{21}}{k_{3}^2 \left(n_{2}^2+1\right)+(n_{2} n_{3}+1)^2}+\frac{k_{3}^2 \Delta m^2_{31}}{k_{3}^2+n_{3}^2+1}\right)^{\frac{1}{2}}\hspace{0.9cm}\mbox{for NO},    \\
\sqrt{\frac{\Delta m^2_{21} \left(k_{1} k_{2}+n_{1}^2+1\right)^2}{\left(k_{1}^2+n_{1}^2+1\right) \left[(k_{1} k_{2}+1)^2+\left(k_{2}^2+1\right) n_{1}^2\right]}-\frac{\Delta m^2_{31} \left(n_{1}^2+1\right)}{k_{1}^2+n_{1}^2+1}} \hspace{0.2cm}\,\mbox{for\,  IO},
\end{array}%
\right. \label{mb3v}\\
&& \langle m^{(3)}_{ee}\rangle =\left\{
\begin{array}{l}
\left|\frac{(n_{2} n_{3}+1)^2 \sqrt{\Delta m^2_{21}}}{k_{3}^2 \left(n_{2}^2+1\right)+(n_{2} n_{3}+1)^2}+\frac{k_{3}^2 \sqrt{\Delta m^2_{31}}}{k_{3}^2+n_{3}^2+1}\right|\hspace{1.0 cm}\mbox{for NO},    \\
\hspace{-0.04 cm}\left|\frac{\sqrt{\Delta m^2_{21}-\Delta m^2_{31}} \left(k_{1} k_{2}+n_{1}^2+1\right)^2}{\left(k_{1}^2+n_{1}^2+1\right) \left((k_{1} k_{2}+1)^2+\left(k_{2}^2+1\right) n_{1}^2\right)}+\frac{\sqrt{-\Delta m^2_{31}} k_{2}^2 n_{1}^2}{(k_{1} k_{2}+1)^2+k_{2}^2 n _{1}^2+n_{1}^2}\right| \hspace{0.1cm}\,\mbox{for\,  IO}.
\end{array}%
\right.\label{mee3v}
\eea

\subsection{3+1 sterile-active neutrino mixing \label{neutrino4}}
The fact that the neutrino mass spectrum can be NO ($m_1< m_2 < m_3$) or IO ($m_3< m_1< m_2$) depending on the sign of $\Delta m^2_{31}$ \cite{Salas2020, Kelly2021}. In MES scenario, the light neutrino masses are given in terms of three neutrino mass-squared differences as
\bea
\left\{
\begin{array}{l}
m_1=0, \hs m_2 =\sqrt{\Delta m_{21}^2}, \hs
m_3=\sqrt{\Delta m_{31}^2}, \hs
m_4=\sqrt{\Delta m_{41}^2}\hspace{0.5cm}\mbox{for NO},    \\
m_3=0, \hs m_2 =\sqrt{\Delta m_{21}^2-\Delta m_{31}^2}, \hs
m_1=\sqrt{-\Delta m_{31}^2}, \hs
m_4=\sqrt{\Delta m_{41}^2-\Delta m_{31}^2} \hspace{0.2cm}\,\mbox{for  IO},
\end{array}%
\right. \label{massorderings}
\eea
where $\Delta m_{ij}^2 =m_j^2-m_i^2 \,\, (ij=21,31,41).$

The mass of sterile neutrino\cite{Barry2011v}, $m_s\simeq -M_{S} M_R^{-1}M_{S}^T$, obtained from Eqs. (\ref{MDRS}) and (\ref{massorderings}) as follows:
\bea
m_4\equiv m_{s}=\frac{(\la_{1}+\la_{2}) g^2}{2 d^2-\la_{1} (\la_{1}+\la_{2})}. 
\label{msm4}
\eea
The $3\times 1$ matrix $R$ which control the strength of active-sterile mixing angles\cite{Barry2011v}, $R = M_D M_R^{-1}M_{S}^T $ $(M_{S} M_R^{-1}M_{S}^T)^{-1}$, with the aid of Eqs. (\ref{MDRS}) and (\ref{massorderings}), reads:
\bea
  R =\left(r_{11} \hs\hs r_{21} \hs\hs r_{31}\right)^T, \label{Rmatrixmod}
\eea
where
\bea
&&r_{11}=\frac{b_{31}}{f}-\frac{d (a_{31}+c_{31})}{f (\lambda_1 + \lambda_2)}, \hs  r_{21}=\frac{b_{23}}{f}-\frac{d (a_{23}+c_{23})}{f (\lambda_1 + \lambda_2)}, \hs r_{31}=\frac{b_{12}}{f}-\frac{d (a_{12}+c_{12})}{f (\lambda_1 + \lambda_2)}. \label{rj1}
\eea
The strength of the active-sterile mixing is determined by\cite{Barry2011v} $\left(U_{e4}  \hs\hs U_{\mu 4} \hs\hs U_{\tau 4}\right)^T=
U^\dagger_L R$.
Combining Eqs. (\ref{Uclep}) and (\ref{Rmatrixmod}) yields:
\bea
&&\left(U_{e4}\hspace{0.5 cm}
U_{\mu 4} \hspace{0.5 cm}
U_{\tau 4}\right)^T=\left( r_{11} \hspace{0.5 cm}
 c_\psi r_{21}-e^{-i \alpha} r_{31} s_\psi \hspace{0.5 cm}
 c_\psi r_{31}+e^{i \alpha} r_{21} s_\psi \right)^T. \label{Uemutau4}
\eea
Three active-sterile neutrino mixing angles are given by:
\bea
&&s^2_{14}= |U_{e4}|^2=r^2_{11}, \crn
&&s^2_{24}=\frac{ |U_{\mu 4}|^2}{1- |U_{e 4}|^2}=\frac{c^2_\psi r_{21}^2-2 c_\psi c_\alpha s_\psi r_{21} r_{31} + s^2_\psi r_{31}^2}{1-r_{11}^2},  \crn
&&s^2_{34}=\frac{ |U_{\tau 4}|^2}{1- |U_{e 4}|^2- |U_{\mu 4}|^2}=\frac{c^2_\psi r_{31}^2+s_{2\psi} c_\alpha r_{21} r_{31}+s^2_\psi r_{21}^2}{1-c^2_\psi r_{21}^2+s_{2\psi} c_\alpha r_{21} r_{31}-r_{11}^2-s^2_\psi r_{31}^2}. \label{s14s24s34sq}
\eea\\
Equations (\ref{rj1}), (\ref{aDij}), (\ref{dg}) and (\ref{s14s24s34sq}) show that the active-sterile neutrino mixing angles depend on the VEVs of scalar fields $H, H^', \chi, \rho, \phi_{\nu}$, the cut-off scale $\Lambda$, Majorana masses $\lambda_{1, 2}$, and the Yukawa like coupling constants in the neutrino sectors.

The effective neutrino masses in 3+1 scheme, $m_{\beta} = \left(\sum^4_{i=1} \left|U_{ei}\right|^2 m_i^2 \right)^{\frac{1}{2}}$ and $\langle m_{ee}\rangle = \left| \sum^4_{i=1} U_{ei}^2 m_i \right|$,  read \cite{betdecay2}: 
\bea
&&m_{\beta} = \left\{
\begin{array}{l}
\left(\frac{(n_{2} n_{3}+1)^2 \Delta m^2_{21}}{k_{3}^2 \left(n_{2}^2+1\right)+(n_{2} n_{3}+1)^2}+\frac{k_{3}^2 \Delta m^2_{31}}{k_{3}^2+n_{3}^2+1}+ m^2_{s} s_{14}^2\right)^{\frac{1}{2}}\hspace{0.15cm}\mbox{for NO},    \\
\hspace{-0.15 cm}\sqrt{\frac{\Delta m^2_{21} \left(k_{1} k_{2}+n_{1}^2+1\right)^2}{\left(k_{1}^2+n_{1}^2+1\right) \left[(k_{1} k_{2}+1)^2+\left(k_{2}^2+1\right) n_{1}^2\right]}-\frac{\Delta m^2_{31} \left(n_{1}^2+1\right)}{k_{1}^2+n_{1}^2+1}+m^2_{s} s_{14}^2} \hspace{0.2cm}\mbox{for  IO},
\end{array}%
\right. \label{mb4v}\\
&& \langle m_{ee}\rangle \hspace{-0.1 cm}=\hspace{-0.1 cm}
\left\{
\begin{array}{l}
 \left|\frac{(n_{2} n_{3}+1)^2 \sqrt{\Delta m^2_{21}}}{k_{3}^2 \left(n_{2}^2+1\right)+(n_{2} n_{3}+1)^2}+\frac{k_{3}^2 \sqrt{\Delta m^2_{31}}}{k_{3}^2+n_{3}^2+1}+ m_{s} s_{14}^2\right|\hspace{0.2cm}\mbox{for NO},    \\
\hspace{-0.1 cm}\left|\frac{\sqrt{\Delta m^2_{21}-\Delta m^2_{31}} \left(k_{1} k_{2}+n_{1}^2+1\right)^2}{\left(k_{1}^2+n_{1}^2+1\right) \left((k_{1} k_{2}+1)^2+\left(k_{2}^2+1\right) n_{1}^2\right)}+\frac{\sqrt{-\Delta m^2_{31}} k_{2}^2 n_{1}^2}{(k_{1} k_{2}+1)^2+k_{2}^2 n _{1}^2+n_{1}^2}+ m_{s} s_{14}^2 \right|  \,\mbox{for  IO}. \hspace*{0.5cm}
\end{array}%
\right.\label{mee4v}
\eea
\section{\label{DMA4} Dark Matter phenomenology}
If sterile neutrino mass $m_s$ is in keV-scale and the active-sterile neutrino mixing angles are tiny, it can be a warm DM candidate \cite{Gariazzojpg15,adhikari2017white,Boyarsky2019,Mertensjpg19,Akerepjc23}. Thus, it is important to calculate the sterile-active mixing to investigate the DM phenomenology. The resulting relic abundance is proportional to the sterile neutrino mass $m_s$ and the sterile-active mixing as follows \cite{ AsakaDM}:
\bea
\Omega_{\mathrm{D}} h^{2} \simeq 3 {\left(\frac{s^{2}_{2\theta}}{10^{-5}}\right)}{\left(\frac{m_{s}}{\mathrm{keV}}\right)}^{2},  \label{Omdm}
\eea
where $s^{2}_{2\theta}$ is the sum of the active-sterile mixing angles,
\bea
s^{2}_{2\theta} = 4\big(|U_{1 4}|^{2}+|U_{2 4}|^{2}+|U_{3 4}|^{2}\big),
\eea
with $U_{i 4}$ are determined in Eq. (\ref{Uemutau4}).

The sterile neutrino is not totally stable and may decay into an active neutrino and a photon $\gamma$ via the process $\nu_s\longrightarrow \nu+\gamma$. However, the decay rate $\Gamma$ is negligible because of the tiny of sterile mixing angles \cite{adhikari2017white}. The decay rate is given by \cite{KennyDM},
\bea
\Gamma\, \big(s^{-1}\big)=1.38\times10^{-32}{\left(\frac{s^{2}_{2\theta}}{10^{-10}}\right)}{\left(\frac{m_{s}}{\mathrm{keV}}\right)}^{5}. \label{Gadm}
\eea
Equations (\ref{Omdm}) and (\ref{Gadm}) show that the relic abundance $\Omega_{\mathrm{DM}} h^{2} $ and the decay rate $\Gamma$ depend on mass and the mixings of sterile neutrino. 
\section{\label{NR} Numerical analysis and discussion}
\emph{We first start with the charged-lepton sector}. Expressions (\ref{h1h2h3h4relation}) and (\ref{del0}) show that at the best-fit values 
\cite{PDG2022}, $m_e=0.51099 \,\mathrm{MeV},  m_\mu = 105.65837\,\mathrm{MeV}, m_\tau = 1776.86 \,\mathrm{MeV}$, and the VEV of the scalars in Eq. (\ref{flavonscale}), $h_1$ depends on one parameter $s_\beta$ while $h_{2}$ and $h_{3}$ depend on two parameters $h_{4}$ and $s_\beta$. In order to satisfy the experimental constraint on the dark matter abundance \cite{Planck2015, Planck2018}, Eqs. (\ref{DMconstrain1}) and (\ref{DMconstrain2}), i.e., $\Omega_{\mathrm{D}} h^2\in (0.117, 0.121)$, the following benchmark points are considered:
\bea
&&y \in (5.850,\,5.851)10^{-2}, \hs  z \in (4.854,\, 4.855)10^{-2}, \crn
&&s_\beta \in (0.348,\, 0.351), \hs c_\alpha \in (0.100,\, 0.800), \hs  h_4 \in(0.700,\, 1.100). \label{benchmarkpoints1}
\eea
With the help of Eq. (\ref{benchmarkpoints1}), the decay rate of the process $\nu_s\longrightarrow \nu+\gamma$ in Eq. (\ref{Gadm}) reaches the ranges,
\bea
\Gamma \big(s^{-1}\big) \in (1.680,\, 1.742)\times10^{-26}. \label{Gadmrange}
\eea
Furthermore,  we obtain the possible ranges of the Yukawa couplings in the charged lepton sector as follows:
\bea
&&|h_1|\in (2.367, \, 2.388) 10^{-1},\hs\, |h_2|\in (2.152, \, 2.224) 10^{-1}, \hs\, |h_3|\in (2.839, \, 3.275) 10^{-1}. \label{h1h2h3constrain}\eea
Furthermore, Eqs.(\ref{abcdl}) and (\ref{psi}) show that at the best-fit values of $m_{e, \mu, \tau}$ 
taken from Ref. \cite{PDG2022} and the VEV of the scalars in Eq. (\ref{flavonscale}), $s_\psi$ depends on two parameters $s_{\beta}$ and $h_4$. With the help of Eq. (\ref{benchmarkpoints1}) we get the following ranges of $s_\psi$:
\bea
&&s_\psi\in (3.158, 5.021)10^{-1}, \hs \mathrm{i.e.}, \hs \psi^{(\circ)} \in (18.410, 30.140). \label{hirange}\eea
\emph{Next, we start with the three active neutrino scheme}. Equations (\ref{k1}) and (\ref{k2}) imply that $\kappa_{1, 2}$ (thus $m_{2,3}$) depend on two observed parameters $\Delta m^2_{21}$ and $\Delta m^2_{31}$. Given $\Delta m^2_{21}$ and $\Delta m^2_{31}$ inside the $3\sigma$ experimentally allowed range\footnote{Numbers are displayed with four significant digits. } taken from Ref. \cite{Salas2020}, i.e., $\Delta m^2_{21}\in (69.4, 81.4)\, \mathrm{meV}^2$ and $\Delta m^2_{31}\in (2.47, 2.63)10^3\, \mathrm{meV}^2$ for NO while $\Delta m^2_{31}\in (-2.53, -2.37)10^3\, \mathrm{meV}^2$ for IO, from Eqs. (\ref{m123}), (\ref{k1}) and (\ref{k2}), we can estimate the values of $\kappa_{1,2}$ and $m_{2,3}$:
\bea
&&\kappa_{1}\in \left\{
\begin{array}{l}
(29.010, 30.150) \,\mathrm{meV} \hspace{0.1 cm} \mbox{for  NO,} \\
(49.040, 50.700)  \,\mathrm{meV} \hspace{0.1 cm}  \mbox{for  IO,}
\end{array}%
\right. \hs 
\kappa_{2}\in \left\{
\begin{array}{l}
(20.340, 21.480) \mathrm{meV} \,\hspace{0.1 cm} \mbox{for  NO,} \\
(0.3426, 0.4145) \mathrm{meV} \, \hspace{0.1 cm}  \mbox{for IO.}%
\end{array}%
\right.  \label{kapa2}\\
&&\left\{
\begin{array}{l}
m_1=0,\hs m_{2}\in(8.331, 9.022)\, \mathrm{meV}, \hspace{0.5 cm} m_{3}\in (49.700, 51.280)\, \mathrm{meV} \,\hspace{0.2 cm} \mbox{for  NO,} \\
m_3=0,\hs m_{1}\in(48.680, 50.300)\, \mathrm{meV}, \,\, m_{2}\in (49.390, 51.100)\, \mathrm{meV} \,\hspace{0.2 cm}  \mbox{for IO.}%
\end{array}%
\right.  \label{mi}\eea
Furthermore, Eqs. (\ref{abcdl}), (\ref{psi}) -
(\ref{del0}), (\ref{kiniexpressionNH}) and (\ref{kiniexpressionIH}) imply that $n_1$ (NO) and $n_3$ (IO) depend on ten parameters including three observed parameters $\theta_{23}, m_\mu, m_\tau$, three scalar VEVs $v_{\mathrm{w}}, v_l, v_\rho$, the cut-off scale $\La$ and three free parameters $h_4, c_\alpha, s_\beta$; $k_1$ (NO) and $k_3$ (IO) depend on eleven parameters including four observed parameters $\theta_{13}, \theta_{23}, m_\mu, m_\tau$, three scalar VEVs $v_{\mathrm{w}}, v_l, v_\rho$, the cut-off scale $\La$ and three free parameters $h_4, c_\alpha, s_\beta$; $n_{2}$ (thus $n_1$ and $k_{1, 2}$) for NO and $k_{2}$ (thus $k_3$ and $n_{2,3}$) for IO depend on twelve parameters including five observed parameters $\theta_{12}, \theta_{13}, \theta_{23}, m_\mu, m_\tau$, three scalar VEVs $v_{\mathrm{w}}, v_l, v_\rho$, the cut-off scale $\La$ and three free parameters $h_4, c_\alpha, s_\beta$.
As consequences, Eqs. (\ref{sdv})$-$ (\ref{mee3v}) imply that $\sin \delta_{CP}$ and the leptonic mixing matrix elements in three neutrino scheme $U_{ij} \, (i,j=1,2,3)$ depend on twelve parameters including five observed parameters $\theta_{12}, \theta_{13}, \theta_{23}, m_\mu, m_\tau$, three scalar VEVs $v_{\mathrm{w}}, v_l, v_\rho$, the cut-off scale $\La$ and three free parameters $h_4, c_\alpha, s_\beta$ while $\langle m_{ee}\rangle$ and $m_\beta$ depend on fourteen parameters including seven observed parameters $\theta_{12}, \theta_{13}, \theta_{23}, m_\mu, m_\tau, \Delta m^2_{21}, \Delta m^2_{31}$, three scalar VEVs $v_{\mathrm{w}}, v_l, v_\rho$, the cut-off scale $\La$ and three free parameters $h_4, c_\alpha, s_\beta$ in which seven observed parameters $\theta_{12}, \theta_{13}, \theta_{23}, m_\mu, m_\tau, \Delta m^2_{21}, \Delta m^2_{31}$ have been measured with quite high accuracy \cite{Salas2020}, the scalar VEVs and the cut-off scale $\La$ are determined in Eqs. (\ref{higgvevs}) and (\ref{flavonscale}) while $c_\alpha$ and $s_\beta$ are constrained parameters in the sense of $c_\alpha, s_\beta \in (-1.0,1.0)$. Therefore, to find the feasible ranges of the model parameters $k_i, n_i \, (i=1,2,3)$ and $h_4, c_\alpha, s_\beta$ as well as the predictive ranges of the experimental parameters $\sin \delta$, $\langle m_{ee} \rangle$ and $m_\beta$, we fix the observable parameters $\theta_{12}, \theta_{13}, \theta_{23}, \Delta m^2_{21}$ and $\Delta m^2_{31}$ at their best-fit values taken from Ref. \cite{Salas2020}.

At the best-fit points of $m_\mu, m_\tau$ and $\theta_{12}, \theta_{13}, \theta_{23}$ \cite{Salas2020}, i.e., $m_\mu = 105.65837\,\mathrm{MeV}$, $m_\tau = 1776.86 \,\mathrm{MeV}$ \cite{PDG2022} and $s^2_{12} = 0.318, \, s^2_{13} = 2.200\times 10^{-2}, \, s^2_{23} = 0.574$ for NH while $s^2_{13} = 2.225\times 10^{-2}, \, s^2_{23} = 0.578$ for IH, with the aid of Eqs. (\ref{higgvevs}), (\ref{flavonscale}) and (\ref{benchmarkpoints1}), equation (\ref{sdv}) yields the following ranges\footnote{The analytical expressions in Eq. (\ref{sdv}) is for both NH and IH, however, the slight difference values of the best-fit points of $s_{13}$ and $s_{23}$ for NO and IO \cite{Salas2020} lead to the slight 
difference of $\sin \delta_{CP}$ for the numerical analysis result, with $\sin \delta_{CP}\in (-0.8749, -0.3641)\,\, \mathrm{, i.e.,}\,\,\, \delta^{(\circ)}_{CP}\in  (299.000, 338.700)$ for IO.} of $\sin \delta_{CP}$,
\bea
&&\sin \delta_{CP}\in (-0.8738, -0.3632)\,\, \mathrm{, i.e.,}\,\,\, \delta^{(\circ)}_{CP}\in  (299.100, 338.700) \hspace{0.2cm}\mbox{for both NO and IO}, \label{sdrange} \eea
which can reach the 3$\sigma$ experimental range taken from Ref. \cite{Salas2020}.

Similarly, at the best-fit points of the observed parameters and with the help of Eqs. (\ref{higgvevs}) and (\ref{flavonscale}), we can estimate the values of the model parameters $k_i, n_i\, (i=1,2,3)$ and the ranges of the magnitudes of the elements of the lepton mixing matrix $|(U_L)_{ij}|$ as follows:
\bea
&&k_{1}\in \left\{
\begin{array}{l}
(-2.4670, -1.6210) \,  \hspace{0.1cm}\mbox{for  NO},  \\
(0.1639, 0.2302) \,\hspace{0.7cm}\mbox{for IO},
\end{array}%
\right. \hspace{0.35 cm} k_{2}\in \left\{
\begin{array}{l}
(0.9888, 2.3030) \,  \hspace{0.25cm}\mbox{for  NO},  \\
(2.3650, 7.6220) \,\hspace{0.25cm}\mbox{for IO},
\end{array}%
\right. \crn
&&k_{3}\in \left\{
\begin{array}{l}
(0.1626, 0.2267) \,  \hspace{0.675cm}\mbox{for  NO},  \\
(-1.3930, -0.8213) \,\hspace{0.1cm}\mbox{for IO},
\end{array}%
\right. \hspace{0.35 cm} n_{1}\in \left\{
\begin{array}{l}
(0.5580, 1.4280) \,  \hspace{0.25cm}\mbox{for  NO},  \\
(-1.1530, -0.4249) \hspace{0.1cm}\mbox{for IO},
\end{array}%
\right. \crn
&&n_{2}\in \left\{
\begin{array}{l}
(1.0800, 3.2790) \,  \hspace{0.25cm}\mbox{for  NO},  \\
(1.3400, 5.2940) \,\hspace{0.25cm}\mbox{for IO},
\end{array}%
\right. \hspace{0.75 cm} n_{3}\in \left\{
\begin{array}{l}
(-1.1340, -0.4192) \hspace{0.1cm}\mbox{for  NO},  \\
(0.7034, 1.8160) \,\hspace{0.35cm}\mbox{for IO},
\end{array}%
\right. \label{kinirange}
\eea

The effective neutrino masses in three neutrino scheme reach the following ranges:
\bea
&&\langle m^{(3)}_{ee}\rangle \in \left\{
\begin{array}{l}
 (3.684,\, 3.934) \, \mathrm{meV}  \hspace{0.475cm}\mbox{for  NO},  \\
 (47.820,\, 49.430)\, \mathrm{meV} \hspace{0.1cm}\mbox{for IO},
\end{array}%
\right. \label{mee3constraint}\\
&&m^{(3)}_{\beta} \in \left\{
\begin{array}{l}
(8.714,\, 9.120)\, \mathrm{meV}  \hspace{0.575cm}\mbox{for  NO},  \\
(48.360,\, 49.990) \, \mathrm{meV} \hspace{0.2cm}\mbox{for IO}.
\end{array}%
\right. \label{mb3constraint}\eea
The obtained effective neutrino masses for the three neutrino scheme in Eqs. (\ref{mee3constraint}) and (\ref{mb3constraint}) are below all the upper limits taken from CUORE \cite{CUORE20} $\langle m_{ee} \rangle < (75 \div 350) \,\mathrm{meV}$, CUPID-Mo Collaboration \cite{CUPID2021} $\langle m_{ee} \rangle < (310 \div 540) \,\mathrm{meV}$,  MAJORANA Collaboration \cite{Majoranaco2023}\, $\langle m_{ee} \rangle < (113\div 269) \,\mathrm{meV}$ and KamLAND-Zen \cite{KamLAND2023} $\langle m_{ee} \rangle < (36\div 156) \,\mathrm{meV}$.

\emph{Finally, we refer to the $3+1$ neutrino scheme.}
Expressions (\ref{aDij}), (\ref{dg}) and (\ref{msm4}) imply that $m_s$ depends on nine parameters including four scalar VEVs $v_{1,2,3\nu}, v_\chi$, the cut-off scale $\Lambda$, two Majorana masses $\lambda_{1, 2}$, and two Yukawa like couplings $y$ and $z$. Since the right-handed neutrino mass is in the scale of $M_R> 10^9\, \mathrm{GeV}$ \cite{Mrscale}, we can assume that
\bea
&&\lambda_{1}=10^{11}\, \mathrm{GeV}, \hs \lambda_{2}=2\times 10^{11}\, \mathrm{GeV}. \label{l1l2}
\eea
With the help of Eqs. (\ref{flavonscale}) and (\ref{l1l2}), the sterile neutrino mass $m_s$ depends on two Yukawa like couplings $y$ and $z$, $m_s=9375z^2/(2704 y^2-6)$; thus, we find the possible range of $m_s$ as follows
\bea
m_s \in (6.783,\, 6.791)\, \mathrm{keV}, \label{msrange}
\eea
provided that y and z are given in Eq. (\ref{benchmarkpoints1}).

Furthermore, Eqs. (\ref{aDij}), (\ref{dg}), (\ref{rj1}) show that the parameters which determine the strength of active-sterile mixing angles $r_{i1}\, (i=1,2,3)$ depend on eighteen parameters including $v_\chi, v_\rho, v_{1,2,3\nu},  v_{\mathrm{w}}$, $\lambda_{1, 2}$,
$\Lambda, x_{j}\, (j=1\div6), y, z$ and $s_\beta$. It will be convenient to consider the Yukawa couplings in the same scale of magnitude, thus, we will use the following benchmark points:
\bea
&&x_1 = 5\times 10^{-2}, \hs
x_2 = 6\times 10^{-2}, \hs
x_3 = 7\times 10^{-2}, \crn
&&x_4 =10^{-2}, \hs x_5 =2\times 10^{-2}, \hs x_6 =3\times 10^{-2}. \label{xiasum}
\eea
Using Eqs. Eqs. (\ref{higgvevs}), (\ref{flavonscale}), (\ref{benchmarkpoints1}), (\ref{l1l2}) and (\ref{xiasum}), we get the following ranges for $r_{i1}$,
\bea
&&r_{11}\in \big(-3.684,\,-3.608\big)10^{-5}, \hs r_{21} \in \big(-2.582,\, -2.529\big)10^{-5},\crn
&&r_{31} \in \big(-1.318,\, -1.291\big)10^{-5}.  \label{rijrange}
\eea
Next, expressions (\ref{psi}), (\ref{aDij}), (\ref{dg}), (\ref{rj1}) and (\ref{s14s24s34sq}), with the aid of Eqs. (\ref{higgvevs}), (\ref{flavonscale}) and (\ref{l1l2}), imply that $s^2_{14}$ remain depends on three parameters $y, z$ and $s_\beta$ while $s^2_{24}$ and $s^2_{34}$ depend on five parameters $y, z, c_\alpha, s_\beta$ and $h_4$. With the parameter ranges given in Eq. (\ref{benchmarkpoints1}), we get the following constraints:
 \bea
&&|U_{14}|^2\simeq s^2_{14} \in \big(1.304,\, 1.357\big)10^{-9}, \hs
|U_{24}|^2\simeq s^2_{24} \in \big(3.009,\, 5.845\big)10^{-10},\crn
&&|U_{34}|^2\simeq s^2_{34} \in \big(2.408,\, 5.234\big)10^{-10}, \label{si4sq}
\eea
which are small enough to reach the relic abundance and the decay rate of dark matter. It is interesting to note that the obtained results in Eqs. (\ref{msrange}) and (\ref{si4sq}) are in good agreement with the astrophysical X-ray constraints \cite{Schneider16}, the deep Chandra X-ray observation \cite{Hofmann21} as well as the constraints on (keV) sterile neutrino DM \cite{Boyarsky12, Boyarsky14, Dasgupta21} which could be verified by the HUNTER Collaboration \cite{Martoff2021} and KATRIN Collaboration \cite{Mertensjpg19,Benso19}.

The effective neutrino masses, taking into account the contribution of sterile neutrino, reach the following ranges:
\bea
&&\langle m_{ee}\rangle \in \left\{
\begin{array}{l}
\big(3.693, 3.943\big)\, \mathrm{meV}  \hspace{0.575cm}\mbox{for  NO},  \\
\big(47.830, 49.430\big) \, \mathrm{meV} \hspace{0.2cm}\mbox{for IO},
\end{array}%
\right. \label{meeconstraint}\\
&&m_{\beta}\in \left\{
\begin{array}{l}
 \big(245.100,\, 252.200\big)\, \mathrm{meV}  \hspace{0.2cm}\mbox{for  NO},  \\
 \big(249.700,\, 254.900\big)\, \mathrm{meV}  \hspace{0.2cm}\mbox{for IO}.
\end{array}%
\right. \label{mbconstraint}\eea
It is nice to note that the obtained effective neutrino masses 
in Eqs. (\ref{meeconstraint}) and (\ref{mbconstraint}) are below all the upper limits taken from CRES \cite{PJ82023} with $m_\beta < 155$ eV (Bayesian analysis) and $m_\beta < 152$ eV (Frequentist analysis), NEXT collaboration \cite{NEXT2023} with $\langle m_{ee}\rangle < (0.74\div 3.19)$  eV (model-dependent analysis) and $\langle m_{ee}\rangle < (0.48\div 2.07)$  eV (subtraction $\beta\beta$ analysis).
\section{\label{conclusion} Conclusions}
We have proposed a Standard model extension based on the discrete symmetry $A_4\times Z_4\times Z_2\times Z_8$ which successfully explains the recent observed pattern of lepton masses and mixing angles for both three active neutrino scheme and $3+1$ scheme as well as the keV sterile neutrino dark matter in the frame work of the minimal extended seesaw.

\begin{itemize}
\item [$\bullet$] The charged-lepton mass hierarchy is naturally achieved with the Yukawa couplings have the magnitude of order $10^{-1}$
    while neutrino mass hierarchy is achieved with the Yukawa couplings have the magnitude of order $10^{-2}$.
\item [$\bullet$] Active neutrino mixing angles can reached the best-fit points with the Dirac CP violation phase $\delta^{(\circ)}_{CP}\in (299.100, 338.700)$ for both NO and  IO.
    \item [$\bullet$] Active-Sterile neutrino mixing angles are predicted to be $|U_{14}|^2\in \big(1.304,\, 1.357\big)10^{-9}, \hs
|U_{24}|^2\in \big(3.009,\, 5.845\big)10^{-10},\hs |U_{34}|^2 \in \big(2.408,\, 5.234\big)10^{-10}$ with a keV-scale of sterile neutrino mass $m_s \in (6.783,\, 6.791)\, \mathrm{keV}$.
\item [$\bullet$] The effective neutrino masses are predicted to be $\langle m_{ee}\rangle \in (3.684,\, 3.934)  \, \mathrm{meV}$ for NO and $\langle m_{ee}\rangle \in  (47.820,\, 49.430)\, \mathrm{meV}$ for IO, and $m_{\beta} \in (8.714,\, 9.120)$ $ \mathrm{meV}$ for NO and $m_{\beta} \in (48.360,\, 49.990) \, \mathrm{meV}$ for IO in the three neutrino scheme,
      $\langle m_{ee}\rangle \in  \big(3.693, 3.943\big)\, \mathrm{meV}$ for NO and $\langle m_{ee}\rangle \in  \big(47.830, 49.430\big)\, \mathrm{meV}$ for IO, and $m_{\beta} \in \big(245.100,\, 252.200\big)\, \mathrm{meV}$ for NO and $m_{\beta} \in \big(249.700,\, 254.900\big)\, \mathrm{meV}$ for IO in 3+1 neutrino scheme.
\item [$\bullet$] The decay rate of the keV sterile neutrino and its relic abundance
are obtained with $\Omega_{\mathrm{D}} h^2\in (0.117, 0.121)$ and $\Gamma \big(s^{-1}\big) \in (1.680,\, 1.742)\times10^{-26}$ which satisfy the requirements of a
sterile neutrino to be DM candidate with for a narrow range of mass.
\end{itemize}

\section*{Acknowledgments}
This research is funded by Vietnam National Foundation for Science and Technology Development (NAFOSTED) under grant number 103.01-2023.45.
\newpage
\appendix
\section{\label{forbidappen}Forbidden Yukawa terms}
\begin{table}[h]
\begin{center}
\vspace{-0.25 cm}
\caption{Forbidden Yukawa terms}
\vspace{-0.35 cm}
 \begin{tabular}{|c|c|c|c|} \hline
Forbidden terms& Forbidden by  \\ \hline
$(\overline{\psi}_L \nu_{s})_{3} (\widetilde{H}\chi^*\rho)_{1},
(\overline{\nu}^C_s\nu_{1,3R})_{1}(\rho^{2}\chi)_{1^\prime},
(\overline{\nu}^C_s\nu_{1,3R})_{1}(\rho^{*2}\chi)_{1^\prime};
(\overline{\nu}^C_s\nu_{s})_{1}(\chi^{2})_{1^{\prime\prime}} $&$A_4$\\\hline
$(\overline{\psi}_{L} l_{1R})_{3} (H^'\phi_\nu\chi)_{3}; (\overline{\psi}_{L} l_{2,3R})_{3} (H^'\phi_\nu)_{3}, (\overline{\psi}_{L} l_{2,3R})_{3} (H\phi_\nu\rho^*)_{3}; (\overline{\psi}_L \nu_{2R})_{3} (\widetilde{H^\prime }\phi_l)_{3}, $ &\multirow{4}{2 cm}{\hspace{0.8 cm}$Z_2$}  \\
$(\overline{\psi}_L \nu_{2R})_{3} (\widetilde{H}\phi^*_l\phi_l)_{3_{s,a}},
(\overline{\psi}_L \nu_{2R})_{3} (\widetilde{H}\phi^2_\nu)_{3_{s,a}},
(\overline{\psi}_L \nu_{2R})_{3} (\widetilde{H}\phi^*_\nu\phi_\nu)_{3_{s,a}},
(\overline{\psi}_L \nu_{2R})_{3} (\widetilde{H}\phi^{*2}_\nu)_{3_{s,a}},$&\\
$(\overline{\psi}_L \nu_{2R})_{3} (\widetilde{H}\phi_{l}\rho)_{3},
(\overline{\psi}_L \nu_{s})_{3} (\widetilde{H^\prime}\phi^*_l\chi)_{3};
(\overline{\nu}^C_{1R} \nu_{2R})_{1} (\phi^*_l\phi^*_\nu \rho^*)_{1},
(\overline{\nu}^C_{2R} \nu_{1R})_{1} (\phi^*_l\phi^*_\nu \rho^*)_{1},$&\\
$(\overline{\nu}^C_{2R} \nu_{3R})_{1} (\phi^*_l\phi^*_\nu \rho^*)_{1},
(\overline{\nu}^C_{3R} \nu_{2R})_{1} (\phi^*_l\phi^*_\nu \rho^*)_{1}$&\\ \hline

$(\overline{\psi}_{L} l_{1R})_{3} (H^'\phi_l\chi^*)_{3}; (\overline{\psi}_{L} l_{2,3R})_{3} (H\phi_l\phi^*_\nu)_{3_{s,a}}, (\overline{\psi}_{L} l_{2,3R})_{3} (H\phi_l\phi_\nu)_{3_{s,a}}; (\overline{\psi}_{L} l_{2,3R})_{3} (H^'\phi^*_l\rho^*)_{3};$&\multirow{10}{2 cm}{\hspace{0.8 cm}$Z_4$}  \\
$(\overline{\psi}_L \psi^C_L)_{1^{\prime}} \widetilde{H}^2; (\overline{\psi}_L \psi^C_L)_{3} (\widetilde{H}^2\phi_\nu)_{3},
(\overline{\psi}_L \psi^C_L)_{3} (\widetilde{H}^2\phi^*_\nu)_{3}; (\overline{\psi}_L \nu_{1,3R})_{3} (\widetilde{H}\phi_\nu)_{3},
(\overline{\psi}_L \nu_{1,3R})_{3} (\widetilde{H}\phi^*_l\phi_l)_{3_{s,a}};$&\\
$(\overline{\psi}_L \nu_{1,3R})_{3} (\widetilde{H}\phi^2_\nu)_{3_{s,a}},
(\overline{\psi}_L \nu_{1,3R})_{3} (\widetilde{H}\phi^*_\nu\phi_\nu)_{3_{s,a}},
(\overline{\psi}_L \nu_{1,3R})_{3} (\widetilde{H}\phi^{*2}_\nu)_{3_{s,a}}; (\overline{\psi}_L \nu_{1,3R})_{3} (\widetilde{H}\phi^*_l\rho^*)_{3};$&\\
$(\overline{\psi}_L \nu_{1,3R})_{3} (\widetilde{H}^'\phi_\nu\rho^*)_{3};
(\overline{\psi}_L \nu_{2R})_{3} (\widetilde{H}\phi^*_\nu)_{3};
(\overline{\psi}_L \nu_{2R})_{3} (\widetilde{H}^'\phi^*_\nu\rho^*)_{3},
(\overline{\psi}_L \nu_{s})_{3} (\widetilde{H^\prime}\phi_\nu\chi^*)_{3},$&\\
$(\overline{\psi}_L \nu_{s})_{3} (\widetilde{H^\prime}\phi^*_\nu\chi^*)_{3};
(\overline{\nu}^C_{1R} \nu_{2R})_{1} (\phi^*_l\phi_l)_{1}, (\overline{\nu}^C_{2R} \nu_{1R})_{1} (\phi^*_l\phi_l)_{1},
(\overline{\nu}^C_{2R} \nu_{3R})_{1} (\phi^*_l\phi_l)_{1},(\overline{\nu}^C_{3R} \nu_{2R})_{1} (\phi^*_l\phi_l)_{1},$&\\
$(\overline{\nu}^C_{1R} \nu_{2R})_{1} (\phi^*_\nu\phi_\nu)_{1},
(\overline{\nu}^C_{2R} \nu_{1R})_{1} (\phi^*_\nu\phi_\nu)_{1},
(\overline{\nu}^C_{2R} \nu_{3R})_{1} (\phi^*_\nu\phi_\nu)_{1},
(\overline{\nu}^C_{3R} \nu_{2R})_{1} (\phi^*_\nu\phi_\nu)_{1},$&\\
$(\overline{\nu}^C_{1R} \nu_{2R})_{1} (\chi^*\chi)_{1},
(\overline{\nu}^C_{2R} \nu_{1R})_{1} (\chi^*\chi)_{1},
(\overline{\nu}^C_{2R} \nu_{3R})_{1} (\chi^*\chi)_{1},
(\overline{\nu}^C_{3R} \nu_{2R})_{1} (\chi^*\chi)_{1}, $&\\
$(\overline{\nu}^C_{1R} \nu_{2R})_{1} (\rho^*\rho)_{1},
(\overline{\nu}^C_{2R} \nu_{1R})_{1} (\rho^*\rho)_{1},
(\overline{\nu}^C_{2R} \nu_{3R})_{1} (\rho^*\rho)_{1},
(\overline{\nu}^C_{3R} \nu_{2R})_{1} (\rho^*\rho)_{1},
(\overline{\nu}^C_s\nu_{1,3R})_{1}(\phi^{2}_l\chi)_{1},$&\\
$(\overline{\nu}^C_s\nu_{1,3R})_{1}(\phi^{*2}_l\chi)_{1};
(\overline{\nu}^C_s\nu_{s})_{1}(\phi^2_l\rho)_{1},
(\overline{\nu}^C_s\nu_{s})_{1}(\phi^{*2}_l\rho)_{1},
(\overline{\nu}^C_s\nu_{s})_{1}(\phi^*_l\phi_l\rho^*)_{1},$&\\
$(\overline{\nu}^C_s\nu_{s})_{1}(\phi^2_\nu\rho^*)_{1},
(\overline{\nu}^C_s\nu_{s})_{1}(\phi^*_\nu\phi_\nu\rho^*)_{1},
(\overline{\nu}^C_s\nu_{s})_{1}(\phi^{*2}_\nu\rho^*)_{1},
(\overline{\nu}^C_s\nu_{s})_{1}(\chi^{*}\chi\rho^*)_{1},
(\overline{\nu}^C_s\nu_{s})_{1}(\rho^{3})_{1}$&\\
 \hline
$(\overline{\psi}_{L} l_{1R})_{3} (H^'\phi_\nu)_{3}, (\overline{\psi}_{L} l_{1R})_{3}(H\phi^2_l)_{3_{s,a}}, (\overline{\psi}_{L} l_{1R})_{3}(H\phi^*_l\phi_l)_{3_{s,a}}, (\overline{\psi}_{L} l_{1R})_{3}(H\phi^{*2}_l)_{3_{s,a}},$&\multirow{12}{2 cm}{\hspace{0.8 cm}$Z_8$}  \\
$(\overline{\psi}_{L} l_{1R})_{3}(H\phi^*_\nu\phi_\nu)_{3_{s,a}}, (\overline{\psi}_{L} l_{1R})_{3}(H\phi^*_l\chi)_{3}, (\overline{\psi}_{L} l_{1R})_{3}(H\phi_l\chi^*)_{3}, (\overline{\psi}_{L} l_{1R})_{3}(H\phi^*_l\chi^*)_{3},$&\\
$(\overline{\psi}_{L} l_{1R})_{3}(H\phi^*_\nu\rho)_{3}, (\overline{\psi}_{L} l_{1R})_{3}(H\phi_\nu\rho^*)_{3}; (\overline{\psi}_{L} l_{2,3R})_{3}(H\phi^*_l)_{3}, (\overline{\psi}_{L} l_{2,3R})_{3}(H^'\phi_l\phi_\nu)_{3_{s,a}},$ & \\
$ (\overline{\psi}_{L} l_{2,3R})_{3}(H^'\phi^*_l\phi_\nu)_{3_{s,a}}; (\overline{\psi}_{L} l_{2,3R})_{3}(H^'\phi^*_l\rho)_{3};
(\overline{\psi}_{L} l_{2,3R})_{3}(H^'\phi_\nu\chi)_{3}, (\overline{\psi}_{L} l_{2,3R})_{3}(H^'\phi_\nu\chi^*)_{3};$&\\
$(\overline{\psi}_L \psi^C_L)_{1} \widetilde{H^\prime}^{2}; (\overline{\psi}_L \nu_{1,3R})_{3} (\widetilde{H^\prime}\phi^{2}_\nu)_{3_{s,a}},  (\overline{\psi}_L \nu_{1,3R})_{3} (\widetilde{H^\prime}\phi^{*2}_\nu)_{3_{s,a}}; (\overline{\psi}_L \nu_{1,3R})_{3} (\widetilde{H}\phi^*_l\rho^*)_{3}; $&\\
$(\overline{\psi}_L \nu_{1,3R})_{3} (\widetilde{H}^'\phi_\nu\rho)_{3};
(\overline{\psi}_L \nu_{2R})_{3} (\widetilde{H}^'\phi^{2}_l)_{3_{s,a}},
(\overline{\psi}_L \nu_{2R})_{3} (\widetilde{H}^'\phi^{*}_l\phi_l)_{3_{s,a}},
(\overline{\psi}_L \nu_{2R})_{3} (\widetilde{H}^'\phi^{*2}_l)_{3_{s,a}},$&\\
$(\overline{\psi}_L \nu_{2R})_{3} (\widetilde{H}^'\phi^{*}_\nu\phi_\nu)_{3_{s,a}};
(\overline{\psi}_L \nu_{2R})_{3} (\widetilde{H}^'\phi_l\chi)_{3},
(\overline{\psi}_L \nu_{2R})_{3} (\widetilde{H}^'\phi^*_l\chi)_{3},
(\overline{\psi}_L \nu_{2R})_{3} (\widetilde{H}^'\phi_l\chi^*)_{3},$&\\
$(\overline{\psi}_L \nu_{2R})_{3} (\widetilde{H}^'\phi^*_l\chi^*)_{3},
(\overline{\psi}_L \nu_{2R})_{3} (\widetilde{H}^'\phi^*_\nu\rho)_{3};
(\overline{\psi}_L \nu_{s})_{3} (\widetilde{H}^'\phi_l)_{3},
(\overline{\psi}_L \nu_{s})_{3} (\widetilde{H}^'\phi^*_l)_{3};$&\\
$(\overline{\psi}_L \nu_{s})_{3} (\widetilde{H}\phi_l\phi_\nu)_{3_{s,a}},
(\overline{\psi}_L \nu_{s})_{3} (\widetilde{H}\phi^*_l\phi_\nu)_{3_{s,a}},
(\overline{\psi}_L \nu_{s})_{3} (\widetilde{H}\phi_l\rho)_{3},
(\overline{\psi}_L \nu_{s})_{3} (\widetilde{H}\phi^*_l\rho)_{3},$&\\
$(\overline{\psi}_L \nu_{s})_{3} (\widetilde{H}\phi_\nu\chi)_{3},
(\overline{\psi}_L \nu_{s})_{3} (\widetilde{H}\phi_\nu\chi^*)_{3};
(\overline{\nu}^C_{1R} \nu_{2R})_{1} (\phi^2_l)_{1},
(\overline{\nu}^C_{2R} \nu_{1R})_{1} (\phi^2_l)_{1},
(\overline{\nu}^C_{2R} \nu_{3R})_{1} (\phi^2_l)_{1},$& \\
$(\overline{\nu}^C_{3R} \nu_{2R})_{1} (\phi^2_l)_{1},
(\overline{\nu}^C_{1R} \nu_{2R})_{1} (\phi^{*2}_l)_{1},
(\overline{\nu}^C_{2R} \nu_{1R})_{1} (\phi^{*2}_l)_{1},
(\overline{\nu}^C_{2R} \nu_{3R})_{1} (\phi^{*2}_l)_{1},
(\overline{\nu}^C_{3R} \nu_{2R})_{1} (\phi^{*2}_l)_{1},$&\\
$(\overline{\nu}^C_{1R} \nu_{2R})_{1} \rho^2,
(\overline{\nu}^C_{2R} \nu_{1R})_{1} \rho^2,
(\overline{\nu}^C_{2R} \nu_{3R})_{1} \rho^2,
(\overline{\nu}^C_{2R} \nu_{2R})_{1} \rho^2,
(\overline{\nu}^C_{1R} \nu_{2R})_{1} (\rho^{*2}_l)_{1},
(\overline{\nu}^C_{2R} \nu_{1R})_{1} (\rho^{*2}_l)_{1},$&\\ \hline
\end{tabular}
\end{center}
\end{table}

\newpage
\bc
Table II. Forbidden interactions (continued)
\ec
\begin{table}[h]
\begin{center}
\vspace{-0.5 cm}
 \begin{tabular}{|c|c|c|c|} \hline
Forbidden terms& Forbidden by  \\ \hline
$(\overline{\nu}^C_{2R} \nu_{3R})_{1} (\rho^{*2}_l)_{1},
(\overline{\nu}^C_{3R} \nu_{2R})_{1} (\rho^{*2}_l)_{1}; (\overline{\nu}^C_s\nu_{1,3R})_{1}(\phi^2_\nu\phi_l)_{1},
(\overline{\nu}^C_s\nu_{1,3R})_{1}(\phi^{*2}_\nu\phi_l)_{1},$&\multirow{9}{2 cm}{\hspace{0.8 cm}$Z_8$}  \\
$(\overline{\nu}^C_s\nu_{1,3R})_{1}(\phi^{2}_\nu\phi^*_l)_{1},
(\overline{\nu}^C_s\nu_{1,3R})_{1}(\phi^{*2}_\nu\phi^*_l)_{1},
(\overline{\nu}^C_s\nu_{1,3R})_{1}(\phi^{2}_\nu\chi)_{1},
(\overline{\nu}^C_s\nu_{1,3R})_{1}(\phi^{*2}_\nu\chi)_{1}; $&\\
$(\overline{\nu}^C_s\nu_{1,3R})_{1}(\phi^{2}_\nu\chi^*)_{1},
(\overline{\nu}^C_s\nu_{1,3R})_{1}(\phi^{*2}_\nu\chi^*)_{1},
(\overline{\nu}^C_s\nu_{1,3R})_{1}(\phi_l\phi_\nu\rho)_{1},
(\overline{\nu}^C_s\nu_{1,3R})_{1}(\phi^*_l\phi_\nu\rho)_{1},$&\\
$(\overline{\nu}^C_s\nu_{1,3R})_{1}(\phi_l\phi_\nu\rho^*)_{1},
(\overline{\nu}^C_s\nu_{1,3R})_{1}(\phi^*_l\phi_\nu\rho^*)_{1},
(\overline{\nu}^C_s\nu_{2R})_{1}(\phi^*_\nu\phi_\nu\phi_l)_{1},
(\overline{\nu}^C_s\nu_{2R})_{1}(\phi^*_\nu\phi_\nu\phi^*_l)_{1},$&\\
$(\overline{\nu}^C_s\nu_{2R})_{1}(\phi^*_l\phi_l\chi)_{1},
(\overline{\nu}^C_s\nu_{2R})_{1}(\phi^*_\nu\phi_\nu\chi)_{1},
(\overline{\nu}^C_s\nu_{2R})_{1}(\phi^*_l\phi_l\chi^*)_{1},
(\overline{\nu}^C_s\nu_{2R})_{1}(\phi^2_l\chi^*)_{1},$&\\
$(\overline{\nu}^C_s\nu_{2R})_{1}(\phi^{*2}_l\chi^*)_{1},
(\overline{\nu}^C_s\nu_{2R})_{1}(\phi^*_\nu\phi_\nu\chi^*)_{1},
(\overline{\nu}^C_s\nu_{2R})_{1}(\phi_l\phi^*_\nu\rho)_{1},
(\overline{\nu}^C_s\nu_{2R})_{1}(\phi^*_l\phi^*_\nu\rho)_{1},$&\\
$(\overline{\nu}^C_s\nu_{2R})_{1}(\phi_l\phi_\nu\rho^*)_{1},
(\overline{\nu}^C_s\nu_{2R})_{1}(\phi^*_l\phi_\nu\rho^*)_{1},
(\overline{\nu}^C_s\nu_{2R})_{1}(\phi^3_l)_{1},
(\overline{\nu}^C_s\nu_{2R})_{1}(\phi^2_l\phi^*_l)_{1},$&\\
$(\overline{\nu}^C_s\nu_{2R})_{1}(\phi^{*2}_l\phi_l)_{1},
(\overline{\nu}^C_s\nu_{2R})_{1}(\phi^{*3}_l)_{1},
(\overline{\nu}^C_s\nu_{2R})_{1}(\chi^{3})_{1};
(\overline{\nu}^C_s\nu_{s})_{1}(\phi^{2}_l)_{1},
(\overline{\nu}^C_s\nu_{s})_{1}(\phi^{*}_l\phi_l)_{1},$&\\
$(\overline{\nu}^C_s\nu_{s})_{1}(\phi^{*2}_l)_{1};
(\overline{\nu}^C_s\nu_{s})_{1}(\phi^{*}_\nu\phi_\nu)_{1},
(\overline{\nu}^C_s\nu_{s})_{1}(\chi^{*}\chi)_{1},
(\overline{\nu}^C_s\nu_{s})_{1}(\rho^{*}\rho)_{1} $&\\ \hline
\end{tabular}
\end{center}
\end{table}
\newpage
\vspace{-0.75 cm}

\section{\label{Scalarpotential} Scalar potential and vacuum stability}

The total scalar potential is given by\footnote{$U(\epsilon_1 \rightarrow \epsilon_2, \varepsilon_1 \rightarrow \varepsilon_2,\cdots) \equiv U(\epsilon_1, \varepsilon_1,\cdots)\!\!\!\mid_{\{\epsilon_1=\epsilon_2,\, \varepsilon_1=\varepsilon_2,\cdots \}}$.}:
\bea
U_{\mathrm{scal}}&=& U(H) +U(H^')+U(\phi_l)+U(\phi_\nu)+U(\chi)+U(\rho)
+ U(H,H^') +U(H,\phi_l) \crn
&+& U(H,\phi_\nu)+U(H,\chi)+ U(H,\rho)
+ U(H^',\phi_l)+U(H^',\phi_\nu)+U(H^',\chi) \crn
&+&U(H^',\rho)+U(\phi_l,\phi_\nu) + U(\phi_l,\chi) +U(\phi_l,\rho) + U(\phi_\nu,\chi)+ U(\phi_\nu,\rho)
+ U(\chi,\rho),\label{Vtotal}
\eea
where\footnote{The following terms $(\phi^*_l \phi_l)_{\underline{1}^{'}}(\phi^*_l \phi_l)_{\underline{1}^{''}}, (\phi^*_l \phi_l)_{\underline{1}^{''}}(\phi^*_l \phi_l)_{\underline{1}^{'}}; (\phi^*_l \phi_l)_{3_s}(\phi^*_l \phi_l)_{3_a}, (\phi^*_l \phi_l)_{3_a}(\phi^*_l \phi_l)_{3_s}, (\phi^*_l \phi_l)_{3_a}(\phi^*_l \phi_l)_{3_a}$; $(\phi^*_\nu \phi_\nu)_{\underline{3}_s} (\phi^*_\nu \phi_\nu)_{\underline{3}_a}, (\phi^*_\nu \phi_\nu)_{\underline{3}_a} (\phi^*_\nu \phi_\nu)_{\underline{3}_s}, (\phi^*_\nu \phi_\nu)_{\underline{3}_a} (\phi^*_\nu \phi_\nu)_{\underline{3}_a}; (\phi^{*}_l \phi_l)_{\underline{1}^'}(\phi^{*}_\nu \phi_\nu)_{\underline{1}^{''}}, (\phi^{*}_l \phi_l)_{\underline{1}^{''}}(\phi^{*}_\nu \phi_\nu)_{\underline{1}^'}; \big[\phi_l(\phi^*_\nu\phi_\nu)_{\underline{3}_a}\big]_{\underline{1}^{''}} \chi$, $\big[\phi^*_l(\phi^*_\nu\phi_\nu)_{\underline{3}_a}\big]_{\underline{1}^{'}} \chi^*$ are invariant under the considered symmetries, however, they are vanished due to the tensor product of $A_4$ group in the T-diagonal basis and the VEV alignment of $\phi_l$ and $\phi_\nu$ in Eq. (\ref{VEV}). On the other hand, other terms with three and four different scalars are not invariant under considered symmetries; thus, they don't contribute to $U_{\mathrm{scalar}}$. Furthermore, for simplicity, we assume that the couplings in the same type of interactions are in the same scale.}
\bea
&&U(H)=\mu^2_H H^\+H +\lambda^H (H^\+H)^2,\hs U(H^')=U(H, H\rightarrow H^'), \\
&&U(\phi_l)=\mu^2_{l} \phi^*_l \phi_l +\lambda^l_1 (\phi^*_l \phi_l)_{\underline{1}} (\phi^*_l \phi_l)_{\underline{1}}
+\lambda^l_2 (\phi^*_l \phi_l)_{3_s}(\phi^*_l \phi_l)_{3_s}, \\
&&U(\phi_\nu)=\mu^2_{\nu} \phi^*_\nu \phi_\nu +\lambda^\nu_1 (\phi^*_\nu \phi_\nu)_{\underline{1}} (\phi^*_\nu \phi_\nu)_{\underline{1}}
+\lambda^\nu_2 (\phi^*_\nu \phi_\nu)_{\underline{1}^'} (\phi^*_\nu \phi_\nu)_{\underline{1}^{''}}
+\lambda^\nu_3 (\phi^*_\nu \phi_\nu)_{\underline{3}_s} (\phi^*_\nu \phi_\nu)_{\underline{3}_s}, \crn
&& U(\chi)=\mu^2_\chi \chi^*\chi +\lambda^\chi (\chi^* \chi)_{\underline{1}} (\chi^* \chi)_{\underline{1}}, \hs
U(\rho)= U(\chi\rightarrow \rho), \\
&&U(H,H^')=\lambda^{HH^'}_{1} (H^\+ H)_{\underline{1}}(H^{'\+} H^')_{\underline{1}}+\lambda^{HH^'}_2 (H^\+ H^')_{\underline{1}^{'}}(H^{'\+} H)_{1^{''}},\\
&& U(H,\phi_l)=\lambda^{H\phi_l}_{1} (H^\+ H)_{\underline{1}}(\phi^*_l \phi_l)_{\underline{1}}
+\lambda^{H\phi_l}_2 (H^\+\phi_l)_{\underline{3}}(\phi^*_l H)_{\underline{3}},\, U(H,\phi_\nu)=U(H,\phi_l\rightarrow \phi_\nu),  \\
&&U(H,\chi)=\lambda^{H\chi}_1 (H^\+ H)_{\underline{1}}(\chi^* \chi)_{\underline{1}}
+\lambda^{H\chi}_2 (H^\+ \chi)_{\underline{1}^{''}}(\chi^* H)_{\underline{1}^{'}}, \\
&&U(H,\rho)=\lambda^{H\rho}_1 (H^\+ H)_{\underline{1}}(\rho^* \rho)_{\underline{1}}
+ \lambda^{H\rho}_2 (H^\+\rho)_{1^'}(\rho^* H)_{1^{''}}, \eea
\bea
&& U(H^',\phi_l)=\lambda^{H^'\phi_l}_{1} (H^{'\+} H^')_{\underline{1}}(\phi^*_l \phi_l)_{\underline{1}}
+\lambda^{H^'\phi_l}_2 (H^{'\+}\phi_l)_{\underline{3}}(\phi^*_l H^')_{\underline{3}},\, U(H^',\phi_\nu)=U(H^',\phi_l\rightarrow \phi_\nu), \hs \\
&&U(H^',\chi)=\lambda^{H^'\chi}_1 (H^{'\+} H^')_{\underline{1}}(\chi^* \chi)_{\underline{1}}
+\lambda^{H^'\chi}_2 (H^{'\+} \chi)_{\underline{1}^{'}}(\chi^* H^')_{\underline{1}^{''}}, \\
&&U(H^', \rho)=\lambda^{H^'\rho}_1 (H^{'\+} H^')_{\underline{1}}(\rho^* \rho)_{\underline{1}}
+ \lambda^{H^'\rho}_2 (H^{'\+}\rho)_{1}(\rho^* H^')_{1}, \\
&&U(\phi_l,\phi_\nu)=\lambda^{\phi_l\phi_\nu}_1 (\phi^{*}_l \phi_l)_{\underline{1}}(\phi^{*}_\nu \phi_\nu)_{\underline{1}}
+ \lambda^{\phi_l\phi_\nu}_2 (\phi^{*}_l \phi_\nu)_{\underline{1}}(\phi^{*}_\nu \phi_l)_{\underline{1}}
+ \lambda^{\phi_l\phi_\nu}_3 (\phi^{*}_l \phi_\nu)_{\underline{1}^'}(\phi^{*}_\nu \phi_l)_{\underline{1}^{''}} \crn
&&\hspace{1.5 cm}+\, \lambda^{\phi_l\phi_\nu}_4 (\phi^{*}_l \phi_\nu)_{\underline{1}^{''}}(\phi^{*}_\nu \phi_l)_{\underline{1}^'}+\lambda^{\phi_l\phi_\nu}_5 (\phi^{*}_l \phi_l)_{\underline{3}_s}(\phi^{*}_\nu \phi_\nu)_{\underline{3}_s}
+\lambda^{\phi_l\phi_\nu}_6 (\phi^{*}_l \phi_\nu)_{\underline{3}_s}(\phi^{*}_\nu \phi_l)_{\underline{3}_s}\crn
&&\hspace{1.5 cm}+\,\lambda^{\phi_l\phi_\nu}_{7} (\phi^{*}_l \phi_\nu)_{\underline{3}_s}(\phi^{*}_\nu \phi_l)_{\underline{3}_a}
+\lambda^{\phi_l\phi_\nu}_{8} (\phi^{*}_l \phi_\nu)_{\underline{3}_a}(\phi^{*}_\nu \phi_l)_{\underline{3}_s}
+\lambda^{\phi_l\phi_\nu}_{9} (\phi^{*}_l \phi_\nu)_{\underline{3}_a}(\phi^{*}_\nu \phi_l)_{\underline{3}_a}, \\
&&U(\phi_l,\chi)=\lambda^{\phi_l\chi}_1 (\phi^{*}_l \phi_l)_{\underline{1}}(\chi^{*} \chi)_{\underline{1}}
+ \lambda^{\phi_l\chi}_2 (\phi^{*}_l \chi)_{\underline{3}}(\chi^{*} \phi_l)_{\underline{3}}, \hs U(\phi_l,\rho)=U(\phi_l,\chi\rightarrow\rho), \\
&&U(\phi_\nu, \chi)=\lambda^{\phi_\nu\chi}_1 (\phi^{*}_\nu \phi_\nu)_{\underline{1}}(\chi^{*} \chi)_{\underline{1}}
+ \lambda^{\phi_\nu\chi}_2 (\phi^{*}_\nu \chi)_{\underline{3}}(\chi^{*} \phi_\nu)_{\underline{3}},\hs U(\phi_\nu,\rho)=U(\phi_\nu,\chi\rightarrow \rho), \\
&& U(\chi,\rho)=\lambda^{\chi\rho}_1 (\chi^{*} \chi)_{\underline{1}}(\rho^{*} \rho)_{\underline{1}}
+\lambda^{\chi\rho}_2 (\chi^{*} \rho)_{\underline{1}^{''}}(\rho^{*} \chi)_{\underline{1}^'}. \label{Uchirho} \eea
Next, we will prove 
that the scalar VEVs 
in Eq. (\ref{VEV}) satisfy the
minimization condition of the scalar potential $U_{\mathrm{scal}}$ 
in Eqs. (\ref{Vtotal})-(\ref{Uchirho}). Considering the case of real VEVs, 
the minimization condition of $U_{\mathrm{scal}}$ becomes:
\bea
\frac{\partial U_{\mathrm{scal}}}{\partial v_\alpha} &=&0,\hs
\frac{\partial^2 U_{\mathrm{scal}}}{\partial v^2_\alpha} >0 \hs \big(v_\alpha=\{v,\, v^',\, v_l, \, v_1, \, v_2, \, v_3,\, v_\chi,\, v_\rho\}\big), \label{conditionv}
\eea
which corresponding to the following condition\footnote{Here we used the following notations: $\la^{\mathbf{x}} = \la^{\mathbf{x}}_1+ \la^{\mathbf{x}}_2$ with $\mathbf{x}=\big\{HH^', H\phi_l, H\phi_\nu, H\chi, H\rho, H^'\phi_l, H^'\phi_\nu, H^'\chi, H^'\rho$, $ \phi_l\chi, \phi_l\rho, \phi_\nu\chi, \phi_\nu\rho, \chi\rho\big\},\, \la^{l} = 9 \la^{l}_1 + 4 \la^{l}_2,\, \ka^{\phi_l\phi_\nu}_1 =   18 \big(\la^{\phi_l\phi_\nu}_1  +  \la^{\phi_l\phi_\nu}_3\big) + 8 \big(\la^{\phi_l\phi_\nu}_2 + \la^{\phi_l\phi_\nu}_6\big),\, \ka^{\phi_l\phi_\nu}_2 =  9 \big(4\la^{\phi_l\phi_\nu}_1 + \la^{\phi_l\phi_\nu}_9\big)+ 4 \big(\la^{\phi_l\phi_\nu}_6- 2 \la^{\phi_l\phi_\nu}_2\big) + 18 \big(\la^{\phi_l\phi_\nu}_4 + \la^{\phi_l\phi_\nu}_5\big)$.}:
\bea
\mu_H^2 + \la^{H\phi_\nu} (v_1^2 + 2 v_2 v_3) + \la^{H\chi} v_\chi^2 + 2 \la^H v^2 +
 \la^{H\phi_l} v_l^2 + \la^{HH^'} v^{'2} + \la^{H\rho} v_\rho^2 =0, \hspace{0.9 cm} &&\label{eq1v} \\
\mu_{H^'}^2 + \la^{H^'\phi_\nu} (v_1^2 + 2 v_2 v_3) + \la^{H^'\chi} v_\chi^2 + \la^{HH^'} v^2 +
 \la^{H^'\phi_l} v_l^2 + 2 \la^{H^'} v^{'2} + \la^{H^'\rho} v_\rho^2=0, \hspace{0.9 cm} &&\label{eq2v} \\
\mu_{l}^2 + \ka^{\phi_l\phi_\nu}_1 v_1^2 + \ka^{\phi_l\phi_\nu}_2 v_2 v_3 + \la^{\phi_l\chi} v_\chi^2 +
 \la^{H\phi_l} v^2 + \frac{2}{9} \la^{l} v_l^2 + \la^{H^'\phi_l} v^{'2} + \la^{\phi_l\rho} v_\rho^2=0,  \hspace{0.9 cm}&&\label{eq3v}\\
\mu_\nu^2 + 2 \la^{\nu}_1 (v_1^2 + 2 v_2 v_3) + \frac{8 \la^{\nu}_3 v_1^3 - 4 \la^{\nu}_3 (v_2^3 + v_3^3) +
  9 \la^{\nu}_2 (v_2^3 + 4 v_1 v_2 v_3 + v_3^3)}{9 v_1}   \hspace{0.9 cm}&&\crn
+ \la^{\phi_\nu\chi} v_\chi^2 + \la^{H\phi_\nu} v^2 + \ka^{\phi_l\phi_\nu}_1 v_l^2 + \la^{H^'\phi_\nu} v^{'2} + \la^{\phi_\nu\rho} v_\rho^2=0, \hspace{0.9 cm}&&\label{eq4v}\\
\mu_\nu^2+\frac{1}{3} \left(v_{2} v_{3} (12 \la^{\nu}_1+3 \la^{\nu}_2+4 \la^{\nu}_3)+6 v_{1}^2 (\la^{\nu}_1+\la^{\nu}_2)+\frac{v_{1} v_{2}^2 (9 \la^{\nu}_2-4 \la^{\nu}_3)}{v_{3}}\right)  \hspace{0.9 cm}&&\crn
+\la^{H\phi_\nu} v^2+\la^{H^'\phi_\nu} v^{'2}+\frac{1}{2}\ka^{\phi_l\phi_\nu}_2 v_{l}^2+\la^{\phi_\nu\chi} v_{\chi}^2+\la^{\phi_\nu\rho} v_{\rho}^2=0, \hspace{0.9 cm}&&\label{eq5v}\\
\mu_\nu^2+\frac{1}{3} \left(v_{2} v_{3} (12 \la^{\nu}_1+3 \la^{\nu}_2+4 \la^{\nu}_3)+6 v_{1}^2 (\la^{\nu}_1+\la^{\nu}_2)+\frac{v_{1} v_{3}^2 (9 \la^{\nu}_2-4 \la^{\nu}_3)}{v_{2}}\right) \hspace{0.9 cm}&&\crn
+\la^{H\phi_\nu} v^2+\la^{H^'\phi_\nu} v^{'2}+\frac{1}{2}\ka^{\phi_l\phi_\nu}_2 v_{l}^2+\la^{\phi_\nu\chi} v_{\chi}^2+\la^{\phi_\nu\rho} v_{\rho}^2=0, \hspace{0.9 cm}&&\label{eq6v}\\
\mu_\chi^2 + \la^{\phi_\nu\chi} (v_1^2 + 2 v_2 v_3) + 2 \la^{\chi} v_\chi^2 + \la^{H\chi} v^2 +
 \la^{\phi_l\chi} v_l^2 + \la^{H^'\chi} v^{'2} + 2 \la^{\chi\rho} v_\rho^2 =0,  \hspace{0.9 cm}&&\label{eq7v}\\
\mu_\rho^2 + \la^{\phi_\nu\rho} (v_1^2 + 2 v_2 v_3) + 2 \la^{\chi\rho} v_\chi^2 +
 \la^{H\rho} v^2 + \la^{\phi_l\rho} v_l^2 + \la^{H^'\rho} v^{'2} + 2 \la^{\rho} v_\rho^2=0,  \hspace{0.9 cm}&&\label{eq8v}\\
\mu_H^2 + \la^{H\phi_\nu} (v_1^2 + 2 v_2 v_3) + \la^{H\chi} v_\chi^2 + 6 \la^{H} v^2 +
 \la^{H\phi_l} v_l^2 + \la^{HH^'} v^{'2} + \la^{H\rho} v_\rho^2 > 0,  \hspace{0.9 cm}&&\label{ieq1}\\
\mu_{H^'}^2 + \la^{H^'\phi_\nu} (v_1^2 + 2 v_2 v_3) + \la^{H^'\chi} v_\chi^2 + \la^{HH^'} v^2 +
 \la^{H^'\phi_l} v_l^2 + 6 \la^{H^'} v^{'2} + \la^{H^'\rho} v_\rho^2>0, \hspace{0.9 cm}&&\label{ieq2}\\
\mu_l^2 + \ka^{\phi_l\phi_\nu}_1 v_1^2 + \ka^{\phi_l\phi_\nu}_2 v_2 v_3 + \la^{\phi_l\chi} v_\chi^2 +
\la^{H\phi_l} v^2 + \frac{2}{3} \la^{l} v_l^2 + \la^{H^'\phi_l} v^{'2} + \la^{\phi_l\rho} v_\rho^2>0,  \hspace{0.9 cm}&&\label{ieq3}\\
\mu_\nu^2 + \frac{2 (9 \la^{\nu}_1 + 4 \la^{\nu}_3) v_1^2}{3} +
 4 (\la^{\nu}_1 + \la^{\nu}_2) v_2 v_3 + \la^{\phi_\nu\chi} v_\chi^2 + \la^{H\phi_\nu} v^2 +
 \ka^{\phi_l\phi_\nu}_1 v_l^2 + \la^{H^'\phi_\nu} v^{'2} + \la^{\phi_\nu\rho} v_\rho^2>0, \hspace{0.9 cm}&&\label{ieq4}\\
\la^{\nu}_2 (6 v_1 v_2 + v_3^2) +
 \frac{4}{3} \big[(3 \la^{\nu}_1 + \la^{\nu}_3) v_3^2-2 \la^{\nu}_3 v_1 v_2\big]>0,  \hspace{0.9 cm}&&\label{ieq5}\\
4 \la^{\nu}_1 v_2^2 + \frac{4}{3} \la^{\nu}_3 (v_2^2 - 2 v_1 v_3) +
 \la^{\nu}_2 (v_2^2 + 6 v_1 v_3)>0, \hspace{0.9 cm} &&\label{ieq6}\\
\mu_\chi^2 + \la^{\phi_\nu\chi} (v_1^2 + 2 v_2 v_3) + 6 \la^{\chi} v_\chi^2 + \la^{H\chi} v^2 +
\la^{\phi_l\chi} v_l^2 + \la^{H^'\chi} v^{'2} + 2 \la^{\chi\rho} v_\rho^2>0,  \hspace{0.9 cm}&&\label{ieq7}\\
\mu_\rho^2 + \la^{\phi_\nu\rho} (v_1^2 + 2 v_2 v_3) + 2 \la^{\chi\rho} v_\chi^2 + \la^{H\rho} v^2 +
 \la^{\phi_l\rho} v_l^2 + \la^{H^'\rho} v^{'2} + 6 \la^{\rho} v_\rho^2>0. \hspace{0.9 cm} &&\label{ieq8}.\eea
The system of Eqs. (\ref{eq1v})-(\ref{eq8v}) yields the following solution:
\bea
&&\mu_H^2 = -\la^{H\phi_\nu} (v_1^2 + 2 v_2 v_3) - \la^{H\chi} v_\chi^2 - 2 \la^{H} v^2 -
  \la^{H\phi_l} v_l^2 - \la^{HH^'} v^{'2} -\la^{H\rho} v_\rho^2, \label{solutionminimaleq1}\\
&&\mu_{H^'}^2 = -\la^{H^'\phi_\nu} (v_1^2 + 2 v_2 v_3) - \la^{H^'\chi} v_\chi^2 -
\la^{HH^'} v^2 - \la^{H^'\phi_l} v_l^2 - 2 \la^{H^'} v^{'2} -  \la^{H^'\rho} v_\rho^2, \\
&&\mu_l^2 = -\frac{1}{2} \ka^{\phi_l\phi_\nu}_2 (v_1^2 + 2 v_2 v_3) -
  \la^{\phi_l\chi} v_\chi^2 - \la^{H^'\phi_l} v^2 - \frac{2}{9}\la^{l} v_l^2 - \la^{H^'\phi_l} v^{'2} -
  \la^{\phi_l\rho} v_\rho^2, \\
&&\mu_\nu^2 = -2 (\la^{\nu}_1 + \la^{\nu}_2) (v_1^2 +2 v_2 v_3) - \la^{\phi_\nu\chi} v_\chi^2 - \la^{H\phi_\nu} v^2 - \frac{1}{2}\ka^{\phi_l\phi_\nu}_2 v_l^2 - \la^{H^'\phi_\nu} v^{'2} -\la^{\phi_\nu\rho} v_\rho^2, \\
&&\mu_\chi^2 = -\la^{\phi_\nu\chi} (v_1^2 + 2 v_2 v_3) -  2 \la^{\chi} v_\chi^2 - \la^{H\chi} v^2 - \la^{\phi_l\chi} v_l^2 - \la^{H^'\chi} v^{'2} -  2 \la^{\chi\rho} v_\rho^2, \\
&&\mu_\rho^2 = -\la^{\phi_\nu\rho} (v_1^2 + 2 v_2 v_3) -  2 \la^{\chi\rho} v_\chi^2 - \la^{H\rho} v^2 - \la^{\phi_l\rho} v_l^2 - \la^{H^'\rho} v^{'2} -  2 \la^{\rho} v_\rho^2, \\
&&\ka^{\phi_l\phi_\nu}_1 = \frac{1}{2}\ka^{\phi_l\phi_\nu}_2, \hs \la^{\phi_\nu}_3 = \frac{9}{4}\la^{\phi_\nu}_2.\label{solutionminimaleq7}
\eea
With the aid of the solutions (\ref{solutionminimaleq1})-(\ref{solutionminimaleq7}), expressions
(\ref{ieq1})-(\ref{ieq8}) yield the following vacuum stability:
\bea
\lambda^{H}>0,\hs \lambda^{H^'}>0, \hs \lambda^{l}>0, \hs \lambda^{\nu}_1 + \lambda^{\nu}_2>0, , \hs \lambda^{\chi}>0,\hs \lambda^{\rho}>0.\label{ineq}\eea
\vspace{-0.75 cm}
\section{\label{Appkappakn12} The explicit expressions of $\kappa_{1,2}, k_{1,2,3}$ and $n_{1,2,3}$}
The explicit expressions of $\kappa_{1,2}, k_{1,2,3}$ and $n_{1,2,3}$ are:
\bea
&&\kappa_{1}=\frac{\kappa_0}{2 \big(\lambda_{2}^2-\lambda_{1}^2\big)}, \hs
\kappa_{2}=\frac{\sqrt{\kappa_0^2+\kappa_0^{'2} \big(\lambda_{2}^2-\lambda_{1}^2\big)}}{2 \big(\lambda_{2}^2-\lambda_{1}^2\big)},\\
&&\kappa_0=\big(a_{12}^2+a_{23}^2+a_{31}^2+c_{12}^2+c_{23}^2+c_{31}^2\big) \lambda_{1}-2\big(a_{12} c_{12}+a_{23} c_{23}+a_{31} c_{31}\big) \lambda_{2}, \\
&&\kappa^'_0=4 \big[a_{31}^2 (c_{12}^2 + c_{23}^2)  + a_{23}^2 (c_{12}^2 + c_{31}^2) +
   a_{12}^2 (c_{23}^2 + c_{31}^2)- 2 a_{12} a_{31} c_{12} c_{31} \crn
&&\hspace{0.55 cm}  -2 a_{23} c_{23} (a_{12} c_{12} + a_{31} c_{31})\big], \\
&&k_1=\frac{a_{23} c_{12}-a_{12} c_{23}}{a_{31} c_{23}-a_{23} c_{31}},\hs
n_1=\frac{a_{31} c_{12}-a_{12} c_{31}}{a_{23} c_{31}-a_{31} c_{23}}, \\
&&k_2=\frac{\big(a_{31} c_{12} - a_{12} c_{31}\big) \sqrt{\delta_{kn}} + \Delta_{k}}{\Delta_{kn}}, \hs
n_2=\frac{(a_{23} c_{12}-a_{12} c_{23})\sqrt{\delta_{kn}}+\Delta_{n}}{\Delta_{kn}}, \\
&&k_3=\frac{ (a_{12} c_{31}-a_{31} c_{12}) \sqrt{\delta_{kn}}+\Delta_{k}}{\Delta_{kn}},\hs
n_3= \frac{(a_{12} c_{23}-a_{23} c_{12})\sqrt{\delta_{kn}} +\Delta_{n}}{\Delta{kn}},\\
&&\delta_{kn}=\big[(c_{12}^2 + c_{23}^2-a_{31}^2)^2  + 2 (a_{31}^2 + c_{12}^2 + c_{23}^2) c_{31}^2
    + 2 a_{23}^2 (a_{31}^2 - c_{12}^2 + c_{23}^2 - c_{31}^2)+a_{23}^4
     + c_{31}^4 \hspace{0.5 cm} \crn
&&\hspace{0.65 cm}+ 8 a_{23} a_{31} c_{23} c_{31}\big] \lambda_1^2 -
 4 \big[a_{12}^3 c_{12}+(a_{23} c_{23} + a_{31} c_{31}) (a_{23}^2 + a_{31}^2 + c_{12}^2 + c_{23}^2 +
    c_{31}^2)\big] \lambda_1 \lambda_2 \crn
&&\hspace{0.65 cm} +
 2 a_{12}^2 \big[(a_{23}^2 + a_{31}^2 + c_{12}^2 - c_{23}^2 - c_{31}^2) \lambda_1^2 -
    2 (a_{23} c_{23} + a_{31} c_{31}) \lambda_1 \lambda_2 + 2 (c_{12}^2 + c_{23}^2 + c_{31}^2) \lambda_2^2\big]\crn
&&\hspace{0.65 cm}+ 4 a_{12} c_{12} \lambda_1 \big[2 (a_{23} c_{23} + a_{31} c_{31}) \lambda_1 - \big(a_{23}^2 + a_{31}^2 + c_{12}^2 +
       c_{23}^2 + c_{31}^2\big) \lambda_2\big] \crn
&&\hspace{0.65 cm}+ 4 (a_{23}^2 + a_{31}^2) (c_{12}^2 + c_{23}^2 + c_{31}^2) \lambda_2^2 + a_{12}^4 \lambda_1^2,\eea
\newpage
\bea
&&\Delta_{kn}=2 \big\{a_{12} c_{12} (c_{23}^2 + c_{31}^2-a_{23}^2 - a_{31}^2) \lambda_1 +
   c_{12}^2 \big[(a_{23}^2 + a_{31}^2) \lambda_2-(a_{23} c_{23} + a_{31} c_{31}) \lambda_1\big] \crn
&&\hspace{0.65 cm}+
   a_{12}^2 \big[(a_{23} c_{23}+ a_{31} c_{31}) \lambda_1 - (c_{23}^2 + c_{31}^2) \lambda_2\big]\big\},\\
&&\Delta_{k}=\big\{ a_{12} \big[2 a_{23} a_{31} c_{23} + (a_{31}^2 -a_{23}^2- c_{12}^2 + c_{23}^2) c_{31} +
    c_{31}^3\big]- a_{12}^3 c_{31} \big\}\lambda_1 \crn
&&\hspace{0.65 cm}+
 c_{12} \big[a_{31} (c_{12}^2 + c_{23}^2 - c_{31}^2-a_{31}^2)-a_{23}^2 a_{31} - 2 a_{23} c_{23} c_{31}\big] \lambda_1 \crn
&&\hspace{0.65 cm}+
 2 \big[(a_{23}^2 + a_{31}^2) c_{12} c_{31}-a_{12} a_{31} (c_{12}^2 + c_{23}^2 + c_{31}^2)\big] \lambda_2 +
  a_{12}^2 c_{12} (a_{31} \lambda_1 + 2 c_{31} \lambda_2),\\
&&\Delta_{n}=\big\{c_{12} \big[a_{23} c_{31}^2-a_{23} (a_{23}^2 + a_{31}^2 - c_{12}^2 + c_{23}^2) - 2 a_{31} c_{23} c_{31}\big]-a_{12}^3 c_{23}\big\}\lambda_1 \crn
&&\hspace{0.65 cm}+
 a_{12} \big[a_{23}^2 c_{23} + 2 a_{23} a_{31} c_{31} +
    c_{23} (c_{23}^2 + c_{31}^2-a_{31}^2 - c_{12}^2)\big] \lambda_1 \crn
&&\hspace{0.65 cm}+
 2 \big[(a_{23}^2 + a_{31}^2) c_{12} c_{23} - a_{12} a_{23} (c_{12}^2 + c_{23}^2 + c_{31}^2)\big] \lambda_2 +
  a_{12}^2 c_{12} (a_{23} \lambda_1 + 2 c_{23} \lambda_2).
\eea

\end{document}